\documentclass[amsmath,amssymb,prb,showpacs]{revtex4}

\usepackage{graphicx}
\usepackage{dcolumn}
\usepackage{bm}

\begin{document}

\title{A self-consistent treatment of non-equilibrium spin torques \\
in magnetic multilayers}

\author{Asya Shpiro, Peter M. Levy}
\affiliation{Department of Physics, New York University, \\
4 Washington Place, New York, NY 10003}
\author{Shufeng Zhang}
\affiliation{Department of Physics and Astronomy, University of
Missouri-Columbia,\\ Columbia, MO 65211}

\date{\today}

\begin{abstract}
It is known that the transfer of spin angular momenta between current
carriers and local moments occurs near the interface of magnetic layers when
their moments are non-collinear. However, to determine the magnitude of the
transfer, one should calculate the spin transport properties far beyond the
interface regions. Based on the spin diffusion equation, we present a
self-consistent approach to evaluate the spin torque for a number of layered
structures. One of the salient features is that the longitudinal and
transverse components of spin accumulations are inter-twined from one layer
to the next, and thus, the spin torque could be significantly amplified with
respect to treatments which concentrate solely on the transport at the
interface due to the presence of the much longer longitudinal spin diffusion
length. We conclude that bare spin currents do not properly estimate the
spin angular momentum transferred between to the magnetic background; the
spin transfer that occurs at interfaces should be self-consistently
determined by embedding it in our globally diffuse transport calculations.
\end{abstract}


\maketitle

\section{\label{sec1} Introduction}

The concept of using a spin polarized current to switch the orientation of a
magnetic layer was developed by Slonczewski and 
Berger~\cite{slon-berger},
and has been followed up by several others~\cite{others}$^-$\cite{StZ2}. 
Recently, 
we 
have
proposed a way to understand this spin transfer torque by adopting the model
we used to understand magnetoresistance for currents perpendicular to the
plane of the layer (CPP)~\cite{zhang}. Namely, two phenomena, CPP
magnetoresistance (MR) and spin torque, originate from the spin
accumulation. The former is primarily associated with the longitudinal spin
accumulation and the latter is governed by a transverse effect. The
distinguishing feature between previous treatments and the one we recently
outlined lies in our focus on the spin transport for the entire CPP
structure rather than for the interface region alone. The specular
scattering of the current at interfaces between magnetic and nonmagnetic
layers that is attendant to ballistic transmission can create spin 
torque~\cite{slon-berger,others}. Here we start at the opposite extreme and
consider the spin torque due to the bulk of the magnetic layers and the
diffuse scattering at interfaces; as is the case for giant magnetoresistance
(GMR) reality is probably a mixture of these extreme positions.

To understand the significance of our approach, we recall the physics of
CPP-MR. The resistance of the entire CPP structure comes from various
sources of scattering. If one only considers scattering at an interface,
one may concentrate on the calculation of the transmission coefficient for
the interface. However, ballistic transmission across an interface is not
the only physics of the transport in magnetic multilayers. The leads, as
well as impurity scattering in the layers, have to be included in the
calculation of the resistance of the entire CPP structure. Therefore, one
should embed the ballistic interface scattering in the framework of the
diffusive scattering from the bulk of the layers, as well as the diffuse
interface scattering, for a better approach to describing the transport.
This is very much the spirit of CPP transport: the spin transport is
described by macroscopic spin diffusion while the detail of the interface
scattering is treated as a boundary condition. The crossover from
ballistic interface scattering to diffusive scattering has been recently
studied in detail~\cite{Qi}. By analogy, we argue that the spin torque
should be calculated by solving the transport equation for the entire
structure. The interface ballistic spin transfer, which is the center of
the previous discussions, may be physically important; however it should
be embedded in a larger picture, i.e., the ballistic transport across an
interface should be connected to the diffusive transport outside the
interfacial region. We will show in this paper that our semi-classical
formalism supplies a natural framework to incorporate this. Indeed Stiles
and Zangwill have done just this; however we make a key assumption, that
is different from theirs. That is, we assume that a component of the spin
current transverse to the magnetization exists in the magnetic
layers.~\cite{StZang} We should emphasize that our transport calculation
does not conflict with the physics of ballistic transport at interfaces.
In fact, we will see below, the ballistic component can be accommodated
within our formalism. 

\begin{figure}
\includegraphics[width=\textwidth]{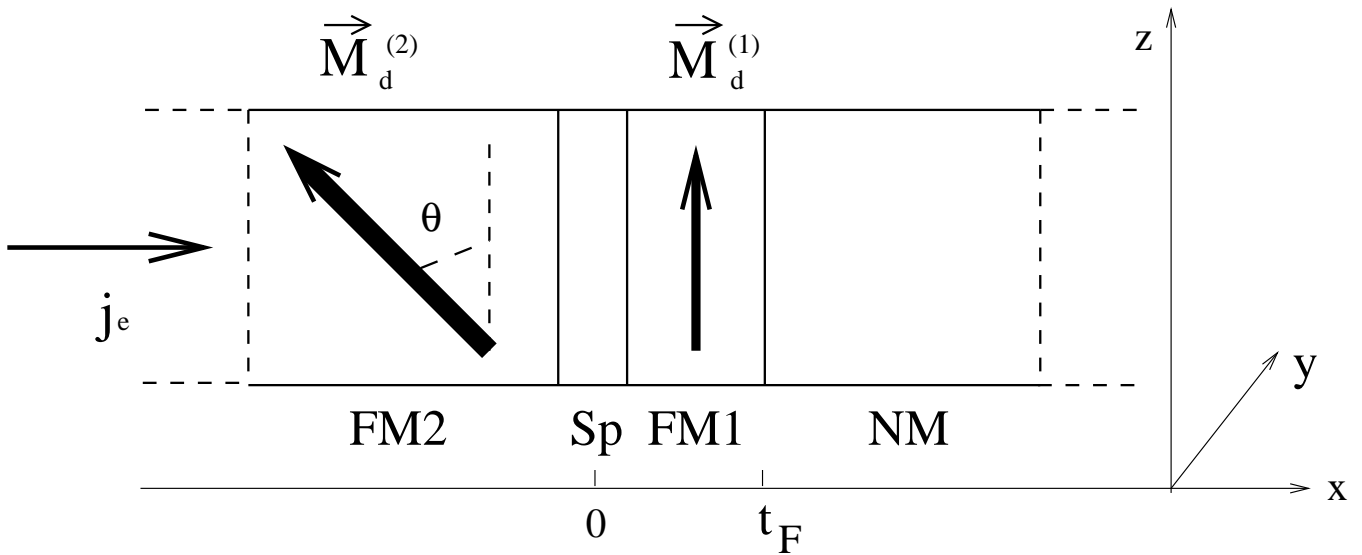}
\caption{Multilayered pillar-like
structure used for current induced reversal of a magnetic layer. FM2 is a
thick ferromagnetic layer with the thickness exceeding $\lambda^F_{sdl}$
and local magnetization ${\bf M}_d^{(2)}=\cos\theta{\bf
e}_z-\sin\theta{\bf e}_y$, Sp is a thin nonmagnetic spacer, FM1 is a thin
ferromagnetic layer with the thickness $t_F$ and local magnetization ${\bf
M}_d^{(1)}={\bf e}_z$, and NM is a nonmagnetic back layer.} 
\label{fig1}
\end{figure}

Here we specify a model system to calculate the spin torque: a magnetic
multilayer whose essential elements consist of a thick magnetic layer,
whose primary role is to polarize the current, a thin magnetic layer that
is to be switched, a nonmagnetic spacer layer so that there is no
interlayer exchange coupling between the thick and thin layers, and a
nonmagnetic layer or lead on back of the thin magnetic layer; see
Fig.~\ref{fig1}.  As we show in this paper the angular momentum
transferred to a thin layer {\it far exceeds} the transverse component (to
the orientation of the magnetization of the thin layer) of the bare
portion of the incoming spin polarized current, i.e., that part
proportional to the electric field; see Eq.~(\ref{c}) below.  It is a
direct consequence of the spin accumulation coming from the two primary
layers, the thick magnetic and nonmagnetic back layers, that produce this
buildup. The role of this accumulation in the spin current is given in
Eq.~(7) of Ref.~\onlinecite{zhang}; it is a consequence of considering the
transport in the multilayer as a diffusive process, and is in keeping with
our previous treatments of transport in magnetic multilayers with the
proviso that one has to include the exchange interaction between the
accumulation and the magnetic background, a.k.a. the ``sd'' interaction,
to obtain the angular momentum transferred. It is this interaction which
produces the spin transfer between the current and the magnetic
background. Among other things the parameters entering our theory are
determined from CPP transport measurements, except for the exchange
interaction between the itinerant electrons and the magnetic background.

In 
this paper we first review the formalism presented in 
Ref.~\onlinecite{zhang} for
calculating the torque and effective field acting on a magnetic layer. In
particular we define the boundary conditions between the layers of the
multilayer and point out what are the sources for the longitudinal and
transverse spin accumulations when the magnetization of the layers are
noncollinear. In Sec.~\ref{sec3} 
we present our results for the spin accumulation,
spin currents and the torque and effective field acting on the thin layer of
the multilayer depicted in Fig.~\ref{fig1}. The bulk of our results are 
obtained
numerically, however in certain limits we are able to give analytic
expressions, e.g., for the amplification of the torque and effective field
acting on a layer due to the accumulation. In Sec.~\ref{sec4} 
we present an analytic
expression for the amplification, and we indicate how recent ``ab-initio'' 
determinations of the change of spin currents across a
ferromagnetic-nonmagnetic (FM/NM) interface can be incorporated at the
interfaces between layers in our diffusive treatment of the transport in a
multilayered structure. The impact of transverse accumulation on GMR is
given in Sec.~\ref{sec5}; 
in particular in creating deviations from the simple linear
dependence of the resistance with the cosine of the angle between the thick
and thin layers. 

\section{\label{sec2} Review of formalism}

By starting with the linear response of the current to the electric field
for diffusive transport~\cite{zhang} 
\begin{equation}
\hat{\jmath}(x)=\hat{C}E(x)-\hat{D}\frac{\partial \hat{n}}{\partial x},
\label{a}
\end{equation}
we find the spin or magnetization current is 
\begin{equation}
{\bf j}_{m}={\rm Re}{\rm Tr}(\mbox{\boldmath $\sigma$}\hat{\jmath})=2{\bf C}
E(x)-2{\bf D}\frac{\partial n_{0}}{\partial x}-2D_{0}\frac{\partial {\bf m}}{
\partial x},  \label{c}
\end{equation}
where both current $\hat{\jmath}(x)$ and field $E(x)$ are directed along the
growth direction $x$ of the multilayer~\cite{aside}. Here $\hat{C}$ 
denotes
a conductivity and $\hat{D}$ a diffusion constant; the two are related by
the Einstein relation $\hat{C}=e^{2}\hat{N}(\epsilon _{F})\hat{D}$ for a
degenerate metal, where $\hat{N}(\epsilon _{F})$ is the density of states at
the Fermi level. The first term on the right hand side of Eq.~(\ref{c}) is 
the
bare contribution to the spin polarized current from the electric field, the
third term is the contribution from the spin accumulation attendant to the
current across a magnetically inhomogeneous structure; although it can be
subsumed as an interface contribution, as we include the diffuse interface
scattering below, we will neglect the second term coming from charge
accumulation. As we will show in magnetic multilayers the contribution of
the third term to the spin current can dominate over the first.

The equation of motion for the spin accumulation is 
\begin{equation}
\frac{\partial {\bf m}}{\partial t}+\frac{{\partial {\bf j}}_{m}}{\partial x}
+(J/\hbar ){\bf m}\times {\bf M}_{d}=-\frac{{\bf m}}{\tau _{sf}},  \label{d}
\end{equation}
where $\tau _{sf}$ is the spin-flip relaxation time of the conduction
electron. While the last term is diffusive the spin motion induced by the
exchange interaction J is not at all diffusive in spin space; it describes a
deterministic or ballistic rotation of the accumulation that itself is
generated by diffusive spin-flip processes.{\em \ }The diffusion equation
for the macroscopic variables ${\bf m}$ and ${\bf j}_{m}$ can be derived
from the Boltzmann equation for the spin distribution function in the limit
where the length scale $d_{J}\equiv v_{F}h/J\gtrsim \lambda _{mfp}$ , i.e.,
when the distance an electron moves while its spin rotates by $2\pi $ is
greater than the mean free path~\cite{gaspari}. In the opposite limit we are
not able to derive such a simple relation between these variables from the
Boltzmann equation; nonetheless we will calculate the spin torque by using
this expression and later evaluate the significance of the results. We
note that in this treatment we consider the effect of the exchange
interaction between the itinerant electrons and the background magnetization
$H_{int}=-J{\bf m}\cdot {\bf M}_{d}$, a.k.a. the ``sd'' interaction, in
the equation of motion for the distribution function. To write the spin
current Eq.~(\ref{c}) in terms of the electric current 
\begin{equation}
j_{e}\equiv {\rm Re}({\rm Tr}\hat{\jmath})=2C_{0}E(x)-2D_{0}\frac{\partial
n_{0}}{\partial x}-2{\bf D}\cdot \frac{\partial {\bf m}}{\partial x},
\label{je}
\end{equation}
we insert this expression in Eq.~(\ref{c}) \ to eliminate the electric 
field
and charge density, and we obtain 
\begin{equation}
{{\bf j}_{m}}=\beta j_{e}{\bf M}_{d}-2D_{0}\left[ \frac{\partial {\bf m}}{
\partial x}-\beta \beta ^{\prime }{\bf M}_{d}({\bf M}_{d}\cdot \frac{
\partial {\bf m}}{\partial x})\right] ,  \label{h}
\end{equation}
where we have dropped an uninteresting term proportional to the derivative
of the charge accumulation $\partial {n_{0}}/\partial x.$ Upon placing this
expression in Eq.~(\ref{d}) we find the equation of motion for the spin
current is 
\begin{widetext}
\begin{equation}
\frac{1}{2D_{0}}\frac{\partial {\bf m}}{\partial t}-\frac{\partial ^{2}{\bf m
}}{\partial x^{2}}+\beta \beta ^{\prime }{\bf M}_{d}\left( {\bf M}_{d}\cdot 
\frac{\partial ^{2}{\bf m}}{\partial x^{2}}\right) +\frac{{\bf m}}{\lambda
_{sf}^{2}}+\frac{{\bf m}\times {\bf M}_{d}}{\lambda _{J}^{2}}=-\frac{1}{
2D_{0}}\frac{\partial }{\partial x}(\beta j_{e}{\bf M}_{d}),  \label{e}
\end{equation}
\end{widetext}
where $\beta ,\beta ^{\prime }$ are the spin polarization parameters defined
by the relations ${\bf C}=\beta C_{0}{\bf M}_{d}$, where ${\bf M}_{d}$ is
the unit vector to represent the direction of the local magnetization and $
{\bf D}=\beta ^{\prime }D_{0}{\bf M}_{d}$~\cite{zhang}, and we have defined $
\lambda _{sf}\equiv \sqrt{2D_{0}\tau _{sf}}$ and $\lambda _{J}\equiv \sqrt{
2\hbar D_{0}/J}=\sqrt{\lambda _{mfp}d_{J}/3\pi }$ where $\tau _{sf}$ is the
spin flip relaxation time, and $J$ the exchange between the itinerant
electrons and the magnetic background. The diffusion constant is $
D_{0}\thicksim (1/3)v_{F}^{2}\tau =(1/3)v_{F}\lambda _{mfp}$ , where $v_{F}$
is the Fermi velocity, $\tau $ \ the momentum relaxation time, and $\lambda
_{mfp}$ the mean free path; $\lambda _{mfp}\equiv v_{F}\tau $. As we
indicate later, if one interprets $\lambda _{mfp}$ as that associated with
the diffusive scattering at interfaces one arrives at a much smaller
estimates for $\lambda _{J}$ than when one uses the mean free path arriving
from scattering in the bulk of the layers.

The term on the right hand side of the time dependent diffusion equation 
Eq.~(\ref{e}) for the spin accumulation is the{\it \ source} term; it is this
term that drives the accumulation~\cite{zhang2}. Here we will look for the
steady state solutions so that the first term on the left hand side is zero,
and the electric current $j_{e}$ is constant throughout the multilayer. \ In
Appendix~\ref{app1} 
we discuss the source term; here we point out that this term
guarantees the continuity of the spin current $j_{m}$\ across the
interfaces, provided\ the accumulation is continuous. This can be
immediately verified by integrating Eq.~(\ref{e}) across an interface 
and using the definition of spin current Eq.~(\ref{h}), or by 
integrating Eq.~(\ref{d}).

We will proceed along the lines of the conventional treatment for current
perpendicular to the plane of the layers (CPP) in magnetic multilayers and
focus on the discontinuous variation of the background magnetization between
the layers. In this treatment of CPP we assume that the magnetization is
uniform throughout a layer so that the source term is confined to interfaces
between layers~\cite{zhang3},~\cite{valet-fert}; in this case one can 
take
into account the source terms by appropriate boundary conditions; this is
the procedure usually followed when calculating the spin accumulation in
magnetic multilayers~\cite{valet-fert}.\ In this case we set the source 
term
in Eq.~(\ref{e}) to zero, separate the spin accumulation into 
longitudinal
(parallel to the {\it local} moment) and transverse (perpendicular to the
{\it \ local} moment) modes, and look for the steady state solutions. We
stress that the terms longitudinal and transverse are relative to the
magnetization in the individual layers, i.e., \ they are {\it locally }
defined and have no global meaning throughout a multilayer. Equation (\ref{e}
) can now be written as 
\begin{equation}
\frac{\partial ^{2}{\bf m}_{||}}{\partial x^{2}}-\frac{{\bf m}_{||}}{\lambda
_{sdl}^{2}}=0,  \label{p}
\end{equation}
where $\lambda _{sdl}=\sqrt{1-\beta \beta ^{\prime }}\lambda _{sf}$, and 
\begin{equation}
\frac{\partial ^{2}{\bf m}_{\perp }}{\partial x^{2}}-\frac{{\bf m}_{\perp }}{
\lambda _{sf}^{2}}-\frac{{\bf m}_{\perp }\times {\bf M}_{d}}{\lambda _{J}^{2}
}=0.  \label{q}
\end{equation}
The longitudinal accumulation ${\bf m}_{||}$ decays at the length scale of
the spin diffusion length $\lambda _{sdl}$ while the transverse spin
accumulation ${\bf m}_{\perp }$ decays as $\lambda _{J}$. It is important to
differentiate between $\lambda _{sdl}$ which arises from spin flip
processes, and $\lambda _{J}$ which represents the decay of transverse spin
currents due to ordinary spin dependent, $\lambda _{mfp}$, but non-spin-flip
scattering. For cobalt we estimate in Appendix~\ref{app1} the transverse spin
accumulation has a much shorter length scale compared to the longitudinal
one; for permalloy $\lambda _{J}$ is comparable to the spin diffusion
length. We limit our present study to $\lambda _{J}\ll \lambda _{sdl}$ .

The boundary conditions are that the spin accumulation ${\bf m}$ and current 
${\bf j}_{m}$ are continuous across at interfaces as long as there is no
specular or diffusive interface scattering; note that it is unnecessary to
invoke the delta function source term at the boundary of a layer (see the
Appendix~\ref{app1}) 
which specifies the discontinuity in $\partial {\bf m/}\partial
x.$ As noted above the source of this discontinuity in the accumulation is
to guarantee the continuity of the spin current (provided no torque is
created at interface, see below), therefore we can simply invoke it; see
Appendix~\ref{app3}. 
While we could in principle consider the effect of specular
scattering at interfaces, as we have for other problems associated with CPP
transport~\cite{shufeng-asya}, this involves considering each ${\bf k}$
vector separately which for a noncollinear multilayer is a cumbersome
problem we will not address. Others have considered specular scattering at
interfaces and have shown it gives rise to spin torque at these 
boundaries~\cite{slon-berger,others}. The presence of diffuse scattering 
at interfaces
due to both roughness and interdiffusion in {\it noncollinear} structures is
treated in Appendix~\ref{app2}.

Once having the spin accumulation we take a look at its influence on the
background magnetization. The equation of motion for the local magnetization
is 
\begin{equation}
\frac{d{\bf M}_{d}}{dt}=-\gamma _{0}{\bf M}_{d}\times ({\bf H}_{e}+J{\bf m}
)+\alpha {\bf M}_{d}\times \frac{d{\bf M}_{d}}{dt},  \label{l}
\end{equation}
where $\gamma _{0}$ is the gyromagnetic ratio, ${\bf H}_{e}$ is the magnetic
field including the contributions from the external field, anisotropy and
magnetostatic field, the additional effective field $J{\bf m}$ is due to
coupling between the local moments (background magnetization) and the spin
accumulation, and the last term is the Gilbert damping term. As seen from
Eq.~(\ref{l}), the longitudinal spin accumulation has no effect on the local
moment; we can re-write Eq.~(\ref{l}) in terms of the transverse spin
accumulation only by replacing ${\bf m}$ by ${\bf m}_{\perp }$. As shown in
Ref. 3 the two components of the accumulation in the plane transverse to the
magnetization $\bf{M}_{d}^{(1)}$ of the layer for which we are calculating the
effect due to the spin current are
\begin{equation}
J\bf{m}_{\perp}=a\bf{M}_{d}^{(2)}\times \bf{M}_{d}^{(1)}+
b\bf{M}_{d}^{(2)}  \label{m}
\end{equation}
where $\bf{M}_{d}^{(2)}$ 
is the magnetization of the other layer; in the case we
discuss in this paper (see Fig.~\ref{fig1}) 
$\bf{M}_{d}^{(1)}$ refers to the thin layer
which is being switched and $\bf{M}_{d}^{(2)}$ 
to the thick layer which polarizes
the current and which is {\it pinned }so that is does not rotate. By placing
this form of the accumulation in the equation of motion for the background
magnetization Eq.~(\ref{l}) we find that the transverse spin 
accumulation produces
two effects simultaneously: the term $b\bf{M}_{d}^{(1)}\times $ 
$\bf{M}_{d}^{(2)}$ is the torque due to an ``effective field''
$b\bf{M}_{d}^{(2)}$,\ and the other is $a\bf{M}_{d}^{(1)}\times 
(\bf{M}_{d}^{(2)}\times
\bf{M}_{d}^{(1)})$ which is called the ``spin torque'' 
predicted by Slonczewski
and Berger~\cite{slon-berger}. The first term produces a precessional 
motion
about $\bf{M}_{d}^{(1)}$; in this sense it acts {\it as if }the spin current
creates a magnetic field on $\bf{M}_{d}^{(1)}$. The second term acts so as to
increase or decrease the angle between $\bf{M}_{d}^{(1)}$ and 
$\bf{M}_{d}^{(2)}$;
also, it acts so as to assist or oppose the damping term in 
Eq.~(\ref{l}).
We stress that it is the ``sd'' exchange interaction between the spin
accumulation attendant to CPP and the background magnetization that are the
origins of the spin torque $a$ and effective field $b$ .

Another way of determining the torque transmitted by the current to the
background comes from recognizing that angular momentum is conserved so that 
${\bf m+M}_{d}=const$, and 
\begin{equation}
{\bf \tau }\thicksim d{\bf m}/dt\equiv - d{\bf M}_{d}/dt.  \label{cc}
\end{equation}
By following this alternate path, which is indeed the path taken by most who
have worked on this problem, we write 
\begin{equation}
{\bf \tau }\thicksim \partial {\bf m}/\partial t+\partial {\bf j}
_{m}/\partial x,  \label{dd}
\end{equation}
which says that the torque transmitted by a steady state current is given by
the gradient of the spin current. When one integrates this over a layer, or
even across an interface, which absorbs the momentum we find

\begin{equation}
\Delta {\bf \tau }=\int_{0}^{t_{F}}(\partial {\bf j}_{m}/\partial x)dx={\bf j
}_{m}(t_{F})-{\bf j}_{m}(0).  \label{ee}
\end{equation}
But from Eqs.~(\ref{d}),~(\ref{h}) and~(\ref{e})

\begin{equation}
(\partial {\bf j}_{m}/\partial x)\thicksim -2D_{0}(\partial ^{2}{\bf m}
/\partial x^{2})\thicksim -2D_{0}\frac{{\bf m\times M}_{d}}{\lambda _{J}^{2}}
.  \label{ff}
\end{equation}
From this form it follows that the $x$ component of the torque comes from
the $y$ component of the accumulation as ${\bf M}_{d}$ is defined as the $z$
direction in the layer which is receiving the spin angular momentum. So
this is contrary what one may think from hastily looking at 
Eq.~(\ref{h})
where one sees that the $x$ component of the spin current is related to the $
x$ component of the gradient of the spin accumulation; however for the
torque it is the {\it gradient} of the spin current that enters. One can
also apply Eq.~(\ref{ee}) across an interface as ab-initio 
calculations~\cite
{others}$^-$\cite{StZ2} have shown that spin dependent specular 
reflections indeed 
induce a
torque at interfaces. In this case Eq.~(\ref{ff}) should be interpreted 
as
the discontinuity of the spin current at an interface; this is concomitant
to either a rotation of the accumulation $m$\ across the interface, a
discontinuity in $m$, or a discontinuity in the background magnetization as
we discuss in Appendix~\ref{app1}.

In the next section we present our results for the spin accumulation and
current, and the ensuing spin torque and field acting on the thin magnetic
layer for the multilayer depicted in Fig.~\ref{fig1}.

\section{\label{sec3} Results}

The essential elements of the multilayered pillar-like structure used for
current induced reversal of a magnetic layer~\cite{myers-fert} are shown 
in Fig.~\ref{fig1}.
The nonmagnetic lead in back of the thick magnetic layer is not
necessary for our discussion as we have taken the thickness of the
ferromagnetic layer to the left  to exceed $\lambda _{sdl}$; what matters is
that the spin polarization of the current is primarily dictated by the thick
magnetic layer. Depending on the thickness of the thin magnetic layer it can
have an effect on the longitudinal accumulation, however for the thicknesses 
$t_{F}$ of interest its role is minor. We have to solve for the spin
accumulation ${\bf m}$ and spin current ${\bf j}_{m}$ in four layers with 3
interfaces: 1) at the thick magnetic layer and the nonmagnetic spacer layer,
2) between the spacer and the thin magnetic layer, and 3) between the thin
layer and the nonmagnetic back layer. At each interface there are 3
components each for ${\bf m}$ and ${\bf j}_{m}$ to match; in all 18
parameters. One approximation, which is valid for all cases of interest, is
to consider the nonmagnetic spacer layer thickness small compared to the
spin diffusion length in the nonmagnetic spacer, i.e., $t_{N}$ $\ll \lambda
_{sdl}^{N}\thicksim 600$ nm. Then both ${\bf m}$ and ${\bf j}_{m}$ are
constant across $t_{N}$ in which case we can focus on three layers and only
12 parameters. In Appendices~\ref{app2} and~\ref{app3} 
we derive the boundary conditions on
the accumulation and current; with these we can determine the spin current
across the entire structure, and consequently the spin torque and effective
field acting on the thin magnetic layer. Without further simplifications we
are unable to give analytic expressions for the accumulation and current
across the multilayer, and present our numerical results for these
quantities as well as the torque and field they create. In the following
section we derive an analytic expression in a limiting case.

\begin{figure*}
\includegraphics[width=\textwidth]{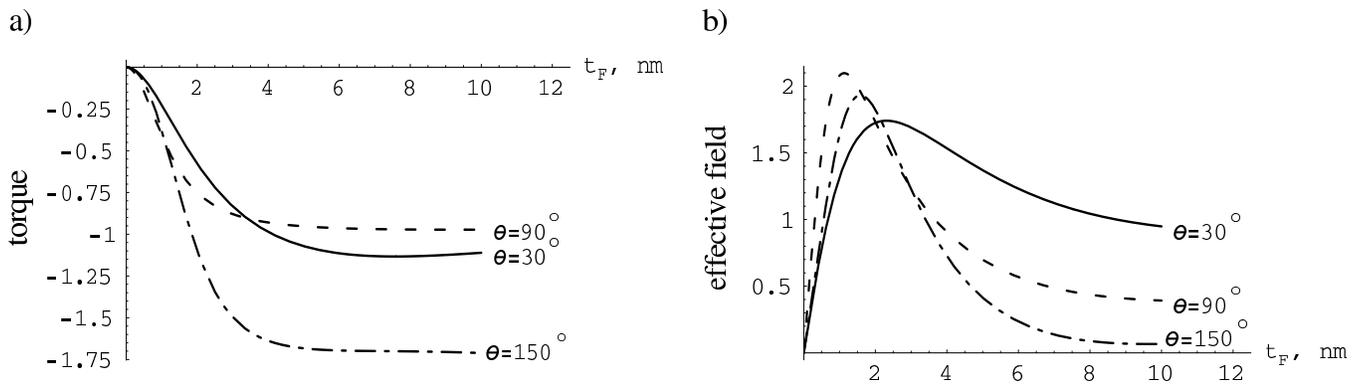}
\caption{Total torque 
$a\sin\theta t_F/\beta j_e\frac{\hbar a_0^3}{e\mu_B}$ (a), 
and total effective field
$b\sin\theta t_F/\beta j_e\frac{\hbar a_0^3}{e\mu_B}$ (b) 
acting on the thin ferromagnetic layer as a
function of its thickness $t_F$ for $\lambda_J=4$ nm,
$\lambda^F_{sdl}=60$ nm, and zero interface resistance $AR_I=0$.}
\label{fig2}
\end{figure*}

In Fig.~\ref{fig2} 
we show the {\it total} (not per unit length as in Ref.~\onlinecite{zhang}) 
spin torque and effective field, see Eq.~(\ref{m}), as a 
function of
the thickness of the thin magnetic layer $t_{F}$ which is being switched;
for these figures we have taken $\lambda _{J}=4$ nm, which is comparable 
to $
\lambda _{mfp}=6$ nm in the bulk, and $\lambda _{sdl}^{F}=60$ nm. While 
this
value of $\lambda _{J}$ may be larger than what one should use for say Co
the plots clearly indicate the new phenomena that occur around the
interfaces. In these plots we have considered neither specular nor
diffuse scattering at the interfaces; the diffusion constant $D_0$ is 
taken to be 
10$^{-3}$ m$^2$/s in the magnetic layers, and 5$\cdot$10$^{-3}$ m$^2$/s in the 
nonmagnetic layers here, as well as in all following plots.
While the torque rapidly increases for
small but finite $t_{F}\approx \lambda _{J}$ and then gradually levels off,
the field is largest about $t_{F}\approx 0.5\lambda _{J}$ and then decreases
towards zero with $t_{F}$; this can be understood as follows. When the
thickness of the thin layer $t_{F}$\ is much smaller than $\lambda _{J}$\
the spin accumulation in the thin layer is the same as that of the thick
layer at the interface, and its direction is parallel to that of the
magnetization of the thick layer $\bf{M}_{d}^{(2)}$ (remember we are not
considering torques created at interfaces); therefore, only the effective
field exists, see Eq.~(\ref{m}). As $t_{F}$\ increases the spin 
accumulation
in the thin magnetic layer rotates away from $\bf{M}_{d}^{(2)}$ and develops a
transverse component 
$\bf{M}_{d}^{(2)}\times \bf{M}_{d}^{(1)}$, i.e., a spin torque
develops. Indeed when $t_{F}\gg \lambda _{J}$, the spin accumulation is
rotated in the thin layer and thus the component of the spin accumulation in
the plane of the magnetizations decreases rapidly, i.e., the effective field
diminishes faster than the torque. As $t_{F}$\ increases further there are
no additional contributions to either the field and torque because they
represent effects that are centered at the interface with the spacer layer
and averaged over the entire thickness of the thin magnetic layer. Although
the effective field is negligible compared to the torque in the limit of
large $t_{F}$ it is noteworthy that at its maximum the field $b$ is at
least as large as $a$. One also notes that while the torque and field terms $
a$ and $b$ are largest for $\theta=180^{0}$ and~$0^{0}$ they do not act on
the background magnetization because $\sin \theta =0$; here $\theta $ is the
angle between the magnetizations of the thick and thin magnetic layers. The
largest effects are found for $\theta \thicksim 150^{0}-170^{0}$.

\begin{figure*}
\includegraphics[width=\textwidth]{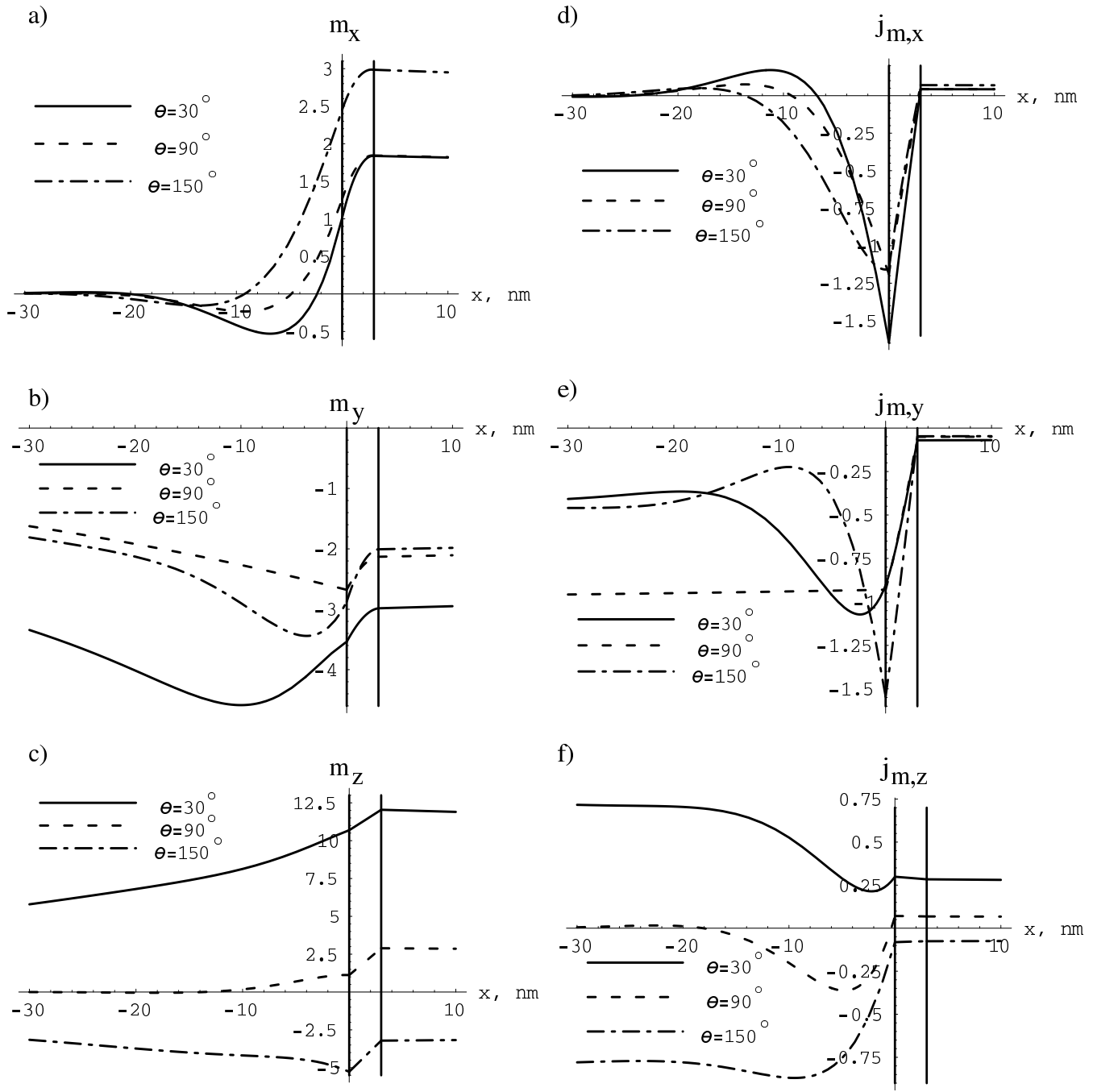}
\caption{$x-$, $y-$, and $z-$components of the spin-accumulation 
${\bf m}/\frac{\beta j_e}{\sqrt 2\lambda_J J}\frac{\hbar a_0^3}{e\mu_B}$ (a-c),
and spin-current ${\bf j}/\beta j_e$  
distribution (d-f) in the structure shown on Fig.1 for
$\lambda_J=4$ nm, $\lambda^F_{sdl}=60$ nm, and zero interface
resistance.}
\label{fig3}
\end{figure*}

In Fig.~\ref{fig3} 
we show the accumulation and spin current that produce these
torques and fields for a thin layer thickness $t_{F}=3$ nm which is
close to where the torque term starts to saturate. As both ${\bf m}$ and $
{\bf j}_{m}$ are constant across $t_{N}$ we do not show the nonmagnetic
spacer layer in these plots, i.e., we plot ${\bf m}$ and ${\bf j}_{m}$ {\it 
as if} $t_{N}=0$. Also we take the magnetization in the thin layer 
$\bf{M}_{d}^{(1)}$ 
as our global $z$ axis; the current is along the $x$ axis which
is along the growth direction of the multilayer, and the $y$ direction is
perpendicular to the other two. In these global axes longitudinal and
transverse in the thin magnetic layer refers to the directions $z$ and $x-y$
; however for the thick layer, whose magnetization $\bf{M}_{d}^{(2)}$ is at an
angle $\theta $ relative to $\bf{M}_{d}^{(1)},$ the global $y$ and $z$ axes do
not define what is meant by longitudinal and transverse in this layer. Also
in the nonmagnetic layers, where there is no equilibrium magnetization, we
talk only about longitudinal accumulation. Far from the interface $x\ll
-\lambda _{J}$ the accumulation and current in the thick layer are collinear
with background magnetization $\bf{M}_{d}^{(2)}$, i.e., referred to its local
axes they are longitudinal with no transverse components, and the spin
current approaches its bare value ${{\bf j}_{m}}\rightarrow \beta j_{e}{\bf M
}_{d}^{(2)}$ (see Eq.~(\ref{h})); even though one still has a 
longitudinal
spin accumulation in the region $-\lambda _{sdl}^{F}\ll x\ll -\lambda _{J}$
its gradient is small compared to that of the transverse accumulation which
makes large contributions to the spin current \ in the region $x>-\lambda
_{J}$. \ This is clear from the plots for $m_{x}$ and $j_{m,x}$ which goes
to zero, while $m_{y}\rightarrow m\sin \theta $, $j_{m,y}\rightarrow \beta
j_{e}\sin \theta ,m_{z}\rightarrow m\cos \theta $, and $j_{m,z}\rightarrow
\beta j_{e}\cos \theta $. With this identification it becomes clear for
example why for $\theta =90^{0}$, $j_{m,y}\rightarrow 1$, in units of $\beta
j_{e}$, while $j_{m,z}\rightarrow 0.$

The results for the spin current in Fig.~\ref{fig3} are interesting; far to 
the left the current is polarized along $\bf{M}_{d}^{(2)}$ 
as one expects in the bulk of a
ferromagnetic. Also in the nonmagnetic layer to the right $x>t_{F}$ there is
no, or very little, spin-current in the $x-y$ directions; the current is
polarized along $\bf{M}_{d}^{(1)},$ i.e., the region $-\lambda _{J}\lesssim
x<t_{F}$ acts as a ``spin filter''~\cite{waintal} inasmuch as the 
component
of the current transverse to the magnetization of the thin layer which is
being switched is ``absorbed'' within this region. That much has been
predicted by most treatments of current induced switching~\cite
{slon-berger,others}. The surprise lies when we look at the enhanced $x-y$
or transverse components of the spin current in the region about $x=0$. As
the ``torques'' (what we call the spin torque and effective field)
transmitted to the thin layer $t_{F}$ are just the difference between the
transverse components of the spin current at the boundaries of the thin
magnetic layer, see Eq.~(\ref{ee}), we find $\tau
_{x}=j_{m,x}(x=t_{F})-j_{m,x}(x=0)\thicksim b$ for the effective field , and 
$\tau _{y}=j_{m,y}(x=t_{F})-j_{m,y}(x=0)\thicksim a$ for the spin torque are
both {\it amplified} when compared to what finds when one neglects spin
accumulation. The thick magnetic layer to the left $x<0$ is pinned so that
the enhanced torques acting in the region of the interface do not produce
any rotation. The $z$ or longitudinal component of the incoming spin current
is not absorbed by the thin magnetic layer as there is no{\em \ }transfer of
spin angular momentum along this direction ($t_{F}\ll \lambda _{sdl}$); see
Fig.~\ref{fig3}f. 
The slight decrease in $j_{m,z}$ is due to the ambient spin flip
scattering in the magnetic layer which is characterized by $\lambda
_{sdl}^{F}\thicksim 60$ nm in the plots shown in Fig.~\ref{fig3}. 
The much slower
decrease in $j_{m,z}(x>t_{F})$ comes from the spin flip scattering in the
nonmagnetic layer whose $\lambda _{sdl}^{N}\thicksim 600$ nm $\gg 
\lambda_{sdl}^{F}$ .

The large enhancement of the transverse spin currents can be understood as
follows. Around $x=0$, which in our picture contains the interfaces between
the thick and thin magnetic layers, the source term for the transverse spin
accumulation is comparable to that for the longitudinal; as mentioned in the
previous section the spin accumulation makes up for the discontinuity in the
``bare'' spin current. At the interface between the thick and thin layers
this is the component of the spin current, coming from the bulk of the thick
magnetic layer, that is perpendicular to the magnetization in the thin
layer. The distance over which it is absorbed is much smaller than that for
the longitudinal accumulation $\lambda _{J}\ll \lambda _{sdl}^{F}$.
Therefore the gradient of the transverse accumulation about $x=0$ is large
and as it is the gradient that contributes to the spin current, 
Eq.~(\ref{h}), we find
the unanticipated amplification of the transverse components of the spin
current at this interface (really two interfaces). This amplification is 
{\it not} a maximum about $\theta =90^{0}$, because in addition to the
''bare'' contribution, there is the component of the spin current
that arises from the longitudinal accumulation in the thick layer, i.e.,
parallel to the magnetization $\bf{M}_{d}^{(2)}$. For $\theta =90^{0}$ this
longitudinal accumulation is quite small, see Fig.~\ref{fig2}h, 
so that there's
little amplification at this angle. In the next section we present a
more quantitative reason for the enhancement.

The plots in Fig.~\ref{fig2} were for the case of $\lambda 
_{sdl}^{F}=60$ nm, $\lambda_{J}=4$ nm, $t_{F}\thicksim$ where the 
torque $a$ starts to saturate. To
determine the roles of the spin diffusion length, $\lambda _{sdl}^{F}$\
(while keeping $\lambda _{sdl}^{N}=600$ nm), the spin transfer length $
\lambda _{J},$\ and the interface resistance due to diffuse scattering at
the interfaces $AR_{I}$ (see Appendix~\ref{app2}), 
on these plots we have rerun our
program for: $\lambda _{sdl}^{F}=30$ nm, $\lambda _{J}=4$ nm, 
$AR_{I}=0$, to show effect of reduced 
$\lambda_{sdl}$ on torques; $\lambda _{sdl}^{F}=60$ nm, $\lambda
_{J}=1$ nm, $AR_{I}=0$, to show effect of smaller $\lambda _{J}$ on 
torques; and $\lambda _{sdl}^{F}=30-60$ nm, $\lambda _{J}=1-4$ nm, 
$AR_{I}\neq 0$, to show effect of interface scattering on torques. We 
account for this scattering in Appendix~\ref{app3} by introducing thin 
``interfacial regions'' in the magnetic
layers which have the enhanced scattering found at interfaces, and derive
the boundary conditions on the accumulation and current in the presence of
interface scattering as we let the thickness of the region tend to zero. The
amount of the scattering $AR_{I}$ and its spin dependence $\gamma $ are
taken from experimental data on CPP-MR~\cite{basspratt}. While the 
diffuse
interface scattering by itself produces sizeable discontinuities in the
accumulation it does not create much spin torque; this is one difference
between our treatment and others.

\begin{figure*}
\includegraphics[width=\textwidth]{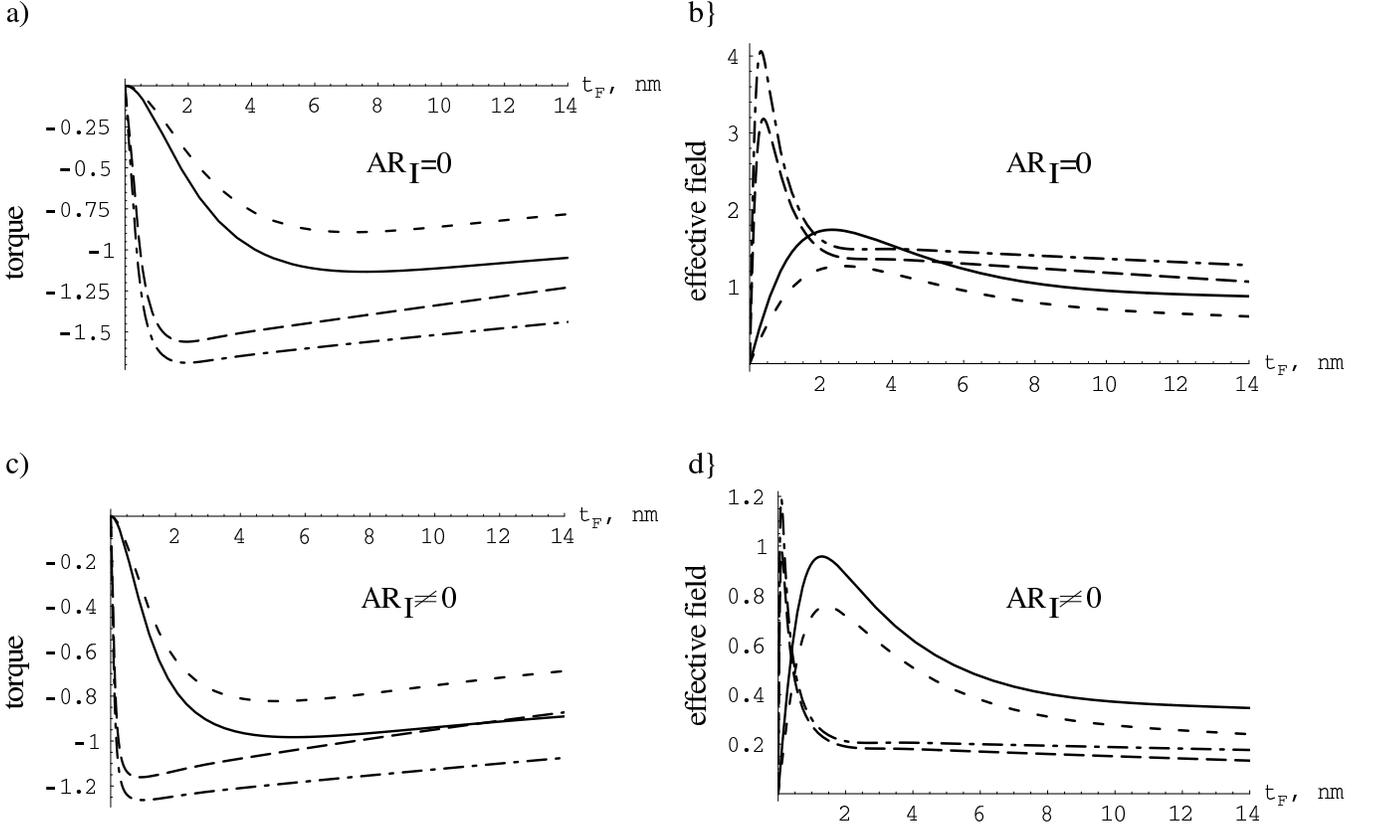}
\caption{Total torque $a\sin\theta t_F/\beta j_e\frac{\hbar a_0^3}{e\mu_B}$
(a,c), and total effective
field $b\sin\theta t_F/\beta j_e\frac{\hbar a_0^3}{e\mu_B}$ 
acting on the thin ferromagnetic layer as a
function of its thickness $t_F$ for $\theta=30^\circ$ for different
values of $\lambda_J$ and $\lambda^F_{sdl}$ in this layer, with and
without interface resistance. Case $\lambda_J=4$ nm, $\lambda_{sdl}=60$ nm 
is represented by the solid line, case $\lambda_J=4$ nm, $\lambda_{sdl}=30$ is 
represented by the short-dashed line, case 
$\lambda_J=1$ nm, $\lambda_{sdl}=60$ by the dashed-dotted line, and case
$\lambda_J=1$ nm, $\lambda_{sdl}=30$ by the long-dashed line.} 
\label{fig4}
\end{figure*}

\begin{figure*}
\includegraphics[width=\textwidth]{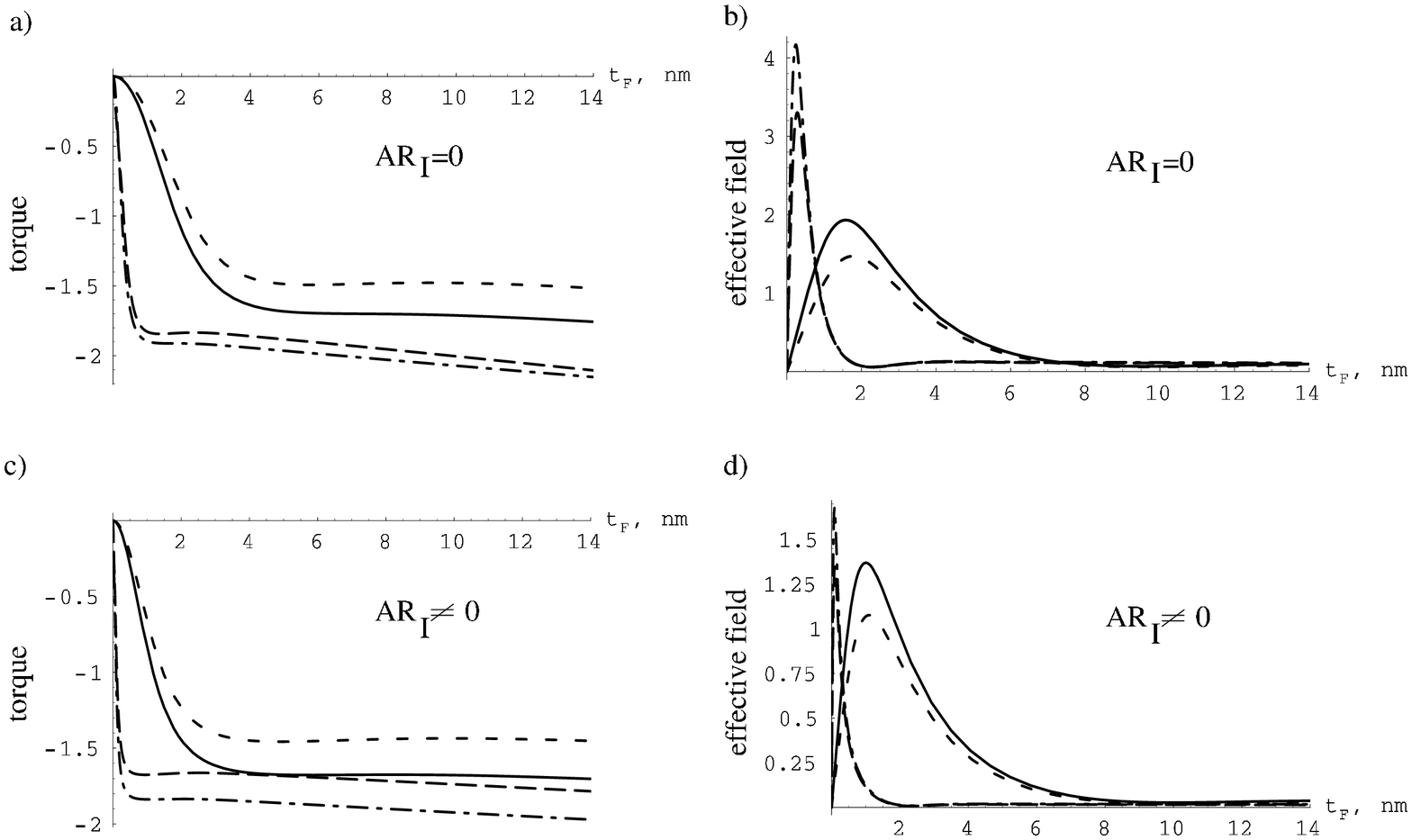}
\caption{Total torque $a\sin\theta t_F/\beta j_e\frac{\hbar a_0^3}{e\mu_B}$ 
(a,c), and total effective
field $b\sin\theta t_F/\beta j_e\frac{\hbar a_0^3}{e\mu_B}$ 
acting on the thin ferromagnetic layer as a
function of its thickness $t_F$ for $\theta=150^\circ$ for different
values of $\lambda_J$ and $\lambda^F_{sdl}$ in this layer, with and
without interface resistance. Case $\lambda_J=4$ nm, $\lambda_{sdl}=60$ nm 
is represented by the solid line, case $\lambda_J=4$ nm, $\lambda_{sdl}=30$ is 
represented by the short-dashed line, case 
$\lambda_J=1$ nm, $\lambda_{sdl}=60$ by the dashed-dotted line, and case
$\lambda_J=1$ nm, $\lambda_{sdl}=30$ by the long-dashed line.}
\label{fig5}
\end{figure*}

In Figs.~\ref{fig4} and~\ref{fig5} 
we summarize our findings by showing plots of the spin
torque $a\sin \theta $\ and effective field $b\sin \theta $\ as functions of
the thickness $t_{F}$ of the thin magnetic layer for different combinations
of $\lambda _{J},\lambda _{sdl}^{F}$, $AR_{I}$, and for two different 
angles $\theta$, 30$^{0}$ and 150$^{0}$, between the magnetic layers. 
We find that
interface resistance increases these torques and causes the spin torque to
achieve saturation for smaller $t_{F}$. By reducing the spin diffusion
length to $\lambda _{sdl}^{F}=30$nm, we find the spin torque and effective
field are reduced. When we reduce the spin transfer length to $\lambda
_{J}=1$ nm we find the spin torque achieves saturation for smaller $t_{F}$
and the effective field is increased and peaks for lower $t_{F}$. 

While the number of plots for the accumulation and currents for
different parameters and thicknesses $t_{F}$ are too numerous to be shown in
this paper, one can view them, in color, at our website 
http://physics.nyu.edu/\~{}avs203 .

\section{\label{sec4} Spin transfer layers}

From the results presented in the last section we arrive at a picture of the
region in which spin angular momentum is transferred that is somewhat
different from the conventional picture~\cite{slon-berger,others}, i.e., 
that
the spin transfer, as it is called, occurs within the few angstroms or
monolayers of the thin magnetic layer that is being switched. In the
conventional picture we would find the transverse spin current in the
ferromagnet would be zero near $x\approx 0$, and the smaller $\lambda _{J}$\
the closer to $x=0.$\ In the thick magnetic layer we find the transverse
components of the spin current goes to zero as expected from 
Eq.~(\ref{q}),
however, for the thin layer we find the region of spin transfer, defined as
the one in which the transverse components of the spin current are absorbed
in Fig.~\ref{fig3}, 
is over the{\it \ entire} thin layer when we take $t_{F}=3$ nm
; this corresponds to $\xi =t_{F}/\sqrt{2}\lambda _{J}=0.5$ for the $\lambda
_{J}=4$ nm we chose. For thinner layers, e.g., $t_{F}=11$\AA , the transverse
spin current $j_{m,y}$ is not entirely absorbed by the layer. To better
understand this behavior we varied the length scale $\lambda _{J}$\ from $
4$ nm to 1 nm, while keeping the $t_{F}=3$ nm the same, i.e., we
varied $\xi $ between $0.5$ and $2$, and found the transverse current still
goes to zero only at $x=t_{F}$. However, for larger thicknesses of $t_{F}$
such that $\xi \gtrsim 5$ and by reducing the resistivity of the nonmagnetic
back layer relative to the magnetic layers (typically by a factor of 5), we
do find the anticipated behavior, i.e., the transverse spin current in the
thin magnetic layer goes to zero before reaching the interface with the
nonmagnetic back layer \ $x=t_{F}$. The reason for the different behavior in
the thin layer arises from its confined geometry, i.e., the reflections from
the thin magnetic layer/nonmagnetic back layer interface create the
patterns observed for the transverse spin currents.

On the basis of our results we can model the transfer of the spin angular
momentum as occurring over a region of several lengths of size $\lambda _{J}$
in the magnetic layers. We arrived at this result by {\it assuming }the
transport in the spin transfer region is diffusive, so that{\em \ }$\lambda
_{J}=\sqrt{\lambda _{mfp}d_{J}/3\pi }$. In a ballistic treatment of this
region one would{\it \ posit} that whatever transverse spin current enters
the thin layer is absorbed by it over a characteristic length $d_{k}\equiv
2\pi \left| k_{\uparrow }-k_{\downarrow }\right| ^{-1}$, i.e., one could
replace our diffusive results with ab-initio calculations of the spin
transfer across the region which is within $d_{k}$ of the 
interface~\cite
{others}$^-$\cite{StZ2}. Having noticed this difference between 
diffusive and 
ballistic
treatments of transport near the interface, it is still necessary to perform
a {\it global} diffusive calculation of the spin accumulation attendant to
CPP across the entire pillar-like structures to determine the actual spin
currents and concomitantly the angular momentum transferred to the
background magnetization; ${\bf j}_{m}$ is not just its bare value $\beta
j_{e}{\bf \hat{M}}_{d}$ in these multilayered structures. Instead, one
should self-consistently determine the transverse {\it and} longitudinal
spin accumulation by taking into account the entire structure as we have
shown in the previous section; here we present a simplified calculation that
gives an analytic expression for the amplification.

Let us consider the limit that the transverse components of the spin
current decay on the length scale $d_{k}\ll t_{F}\ll \lambda _{sf}$.
Based on the band structure and the quantum mechanical probability spin
current $d_{k}$ is only a few angstroms~\cite{others}$^-$\cite{StZ2}. We
can incorporate this picture of interfacial thin ``spin transfer
regions'' into our global diffuse picture of the transport in the same
way as we inserted scattering at interfaces (specular and diffuse) in
our calculation of CPP resistance~\cite {shufeng-asya}. That is, we
posit that the spin accumulation in the region to the left of thin
spacer layer $x<-d,$ where $d=d_{k}$ in the ballistic limit and
$d=\lambda _{J}$ in the diffusive limit, is (see Fig.1) \begin{equation}
{\bf m}(x)=A\hat{{\bf M}}_{2}e^{x/\lambda _{sdl}^{F}}, \label{r}
\end{equation} while for $t_{N}+d<x<t_{N}+t_{F}$ in the thin magnetic
layer that is to be switched, \begin{equation} {\bf m}(x)=(Be^{x/\lambda
_{sdl}^{F}}+Ce^{-x/\lambda _{sdl}^{F}})\hat{{\bf M}
}_{1},  \label{s}
\end{equation}
and in the nonmagnetic back layer for $x>t_{N}+t_{F}$, 
\begin{equation}
{\bf m}(x)=Ee^{-x/\lambda _{sdl}^{N}}\hat{{\bf M}}_{1}.  \label{t}
\end{equation}
The constants are obtained through the boundary conditions as we now show.
These relations explicitly state that the spin accumulation is purely
longitudinal once the electrons are in the ``bulk'' of the ferromagnetic
layers (thick and thin layers) and outside the interfacial spin transfer
regions. To determine the spin torque for each individual ferromagnetic
layer, one needs to find the spin current ${\bf j}_{0}$ and spin
accumulation ${\bf m}_{0}$ in the nonmagnetic spacer layer. By assuming that
the spacer layer $t_{N}$ is small compared to $\lambda _{sdl}^{N\;}$, ${\bf j
}_{0}$ and ${\bf m}_{0}$ are constant across $t_{N}$ $.$ The{\it \ transverse%
} components of ${\bf j}_{0}$ and ${\bf m}_{0}$ vary rapidly in the
interfacial regions at $-d<x<0$ and $t_{N}<x<t_{N}+d$ (indeed they may be
discontinuous), when compared to the {\it longitudinal} spin current and
accumulation which are continuous across the interfaces between the spacer
and the ferromagnetic layers, i.e., are constant across the entire region $
-d<x<t_{N}+d$; therefore the latter provide the bridge between the
accumulations in the regions $x<-d$, Eq.~(\ref{r}) and those for 
$x>t_{N}+d$, Eq.~(\ref{s}).{\em \ }For example, 

\begin{equation}
{\bf m}_{0}\cdot {\bf \hat{M}}_{2}=A,
\end{equation}
and by placing Eq.~(\ref{r}) in Eq.~(\ref{h}) we find 
\begin{equation}
{\bf j}_{0}\cdot {\bf \hat{M}}_{2}=\beta j_{e}-2D_{0}A/\lambda _{sdl}^{F}
\end{equation}
for the interface at $x=0$. Similar expressions can be written down at the $
x=t_{N}$\ and $x=t_{N}+t_{F}$\ interfaces.

As we postulate that the transverse component of the accumulation and
current is limited to a spin transfer region of size $d$\ at the interfaces
we have 
\begin{equation}
\frac{d{\bf m}_{0}}{dx}\times {\bf M}_{1,2}\tilde{=}\frac{{\bf m}_{0}}{d}
\times {\bf M}_{1,2},  \label{u}
\end{equation}
and as the spin current is related to the gradient of the accumulation by
Eq.~(\ref{h}) we find that ${\bf j}_{0}\times {\bf M}_{1,2}\tilde{=}\pm
(D_{0}/d){\bf m}_{0}\times {\bf M}_{1,2}$\ is approximately valid. By 
using this
relation along with the boundary conditions mentioned above, we arrive at
the transverse current density in the spacer layer 
\begin{widetext}
\begin{equation}
{\bf j}_{0}^{\perp }=\beta j_{e}[\sin ^{2}\theta +(2d/\lambda _{sf})\cos
^{2}\theta ]^{-1}{\bf \hat{M}}_{1}\times ({\bf \hat{M}}_{1}\times {\bf \hat{M
}}_{2})  \label{aa}
\end{equation}
\end{widetext}
where we have taken the limit that $d\ll t_{F}\ll \lambda _{sf}$. Clearly,
as $\theta $\ goes to zero or $\pi $, the magnitude of the spin torque is
enhanced by a factor of $\lambda _{sf}/2d$\ compared to the ``bare''
transverse current $\beta j_{e}\hat{{\bf M}}_{1}\times 
(\hat{{\bf M}}_{1}\times 
\hat{{\bf M}}
_{2})$. This huge enhancement comes from interplay between longitudinal and
transverse accumulation; it is the result of the global nature of the spin
current even though the transverse component of the spin current and
accumulation are absorbed within a region $d$ of the interfaces. One should
not take the limit $d\rightarrow 0$ because the assumptions we made about
the spin accumulation, Eqs.~(\ref{r}-\ref{t}) and~(\ref{u}) break down,
i.e., in our calculation the transverse current 
$\bf{j}_{0}^{\perp }$ cannot be
absorbed within $d$ as it tends to zero, and the enhancement does not blow
up.

There is an analogy with how one treats depletion layers in semiconductor
p-n junctions; while the transport is treated by the diffusion equation the
characteristics of the depletion layers themselves are determined from
quantum mechanics. Similarly while the matching of the Boltzmann
distribution functions across interfaces are described by quantum mechanics,
the overall transport in the magnetic multilayered structure is a problem of
diffusive transport.

\section{\label{sec5} Corrections to CPP resistance}

The resistance of a magnetic multilayer for CPP has been extensively
discussed. At first one limited oneself to {\it nominally }collinear
configurations of the magnetic layers, i.e., ferromagnetic and
antiferromagnetic alignments $\theta =0,180^{0}$~\cite{zhang3},~\cite
{valet-fert}, and noncollinear structures were considered where we took
account only of the spin dependent scattering through layer dependent self
energies, but left out the effect of the background magnetization on the
band structure, i.e., we considered conduction by free electron 
states~\cite{camblong}. 
The effect of band structure on CPP resistance has been
considered by Vedyayev~\cite{vedyayev}.

When the idea of current induced switching was proposed it was immediately
recognized that the transfer of angular momentum from the polarized current
would have an effect on the voltage drop across the multilayer being 
studied~\cite{slon}; since that time there have been several calculations of 
the CPP
resistance as a function of the angle between magnetic 
layers~\cite{others}$^-$\cite{StZ2}.
Also there has been experimental data on several multilayered structures
that have confirmed that there are corrections to the simple $\cos
^{2}(\theta /2)$ dependence of the CPP resistance~\cite{dauget}. However 
one
impediment was that for multilayered structures one does not have a good
knowledge of the orientation of the magnetization for the individual layers,
so that one does not have good data on the angular variation of the
resistivity. Recently a study of this was made on an exchange biased spin
valve (ESBV) so as to have a precise determination of the angular dependence
of the CPP-MR~\cite{pratt}. The normalized angular dependence of the
resistance was defined as 
\begin{equation}
R_{norm}=\frac{R(\theta )-R(0)}{R(\pi )-R(0)}  \label{n}
\end{equation}
and the data was fit to 
\begin{equation}
R_{norm}=\frac{1-\cos ^{2}(\theta /2)}{1+\chi \cos ^{2}(\theta /2)}.
\label{o}
\end{equation}

Here we present our calculation of the angular dependence of the CPP
resistance based on the spin currents we find by using the diffusion
equation for the spin accumulation in noncollinear structures; see 
Sec.~\ref{sec2}.
By treating two ``thick'' magnetic layers, i.e., neglecting reflections from
the outer boundaries of the layers, with the magnetizations ${\bf M}
_{d}^{(1)}=\cos (\theta /2){\bf e}_{z}+\sin (\theta /2){\bf e}_{y}$, ${\bf M}
_{d}^{(2)}=\cos (\theta /2){\bf e}_{z}-\sin (\theta /2){\bf e}_{y}$, where $
{\bf x}$ is the direction of the electric current, with a nonmagnetic spacer 
$t_{N}\ll \lambda _{sdl}^{N}$, and zero interface resistance we obtain the
Eq.~(\ref{o}) with 
\begin{equation}
\chi =\frac{1}{\lambda }-1,
\end{equation}
where 
\begin{equation}
\lambda =\frac{(1-\beta \beta ^{\prime })\lambda _{J}}{\sqrt{2}\lambda
_{sdl}^{F}}.
\end{equation}
{\em \ \qquad }It should be stressed that this expression for $\chi $ is
based on the assumption that $\lambda _{J}\ll \lambda _{sdl}^{F}${\em \ . }
For cobalt, $\lambda _{J}$\ is of the order of 3 nm, $\lambda _{sdl}^{F}$\
is about 60 nm~\cite{bass}, taking $\beta $\ to be 0.5~\cite{basspratt}, 
and
calculating $\beta ^{\prime }\approx 0.9$\ by using the densities of states
for up and down electrons~\cite{kuising}{\em , }we estimate $\chi $\ to 
be
about 50. We are unable to compare this to data on systems containing cobalt;
the one system that has been accurately measured has been a series of
Py(t)/Cu(20nm)/Py(t) ESBV with variable permalloy thicknesses ranging from $
6-24$ nm. The best fit yielded $\chi =1.17$; however as $\lambda _{J}\cong
\lambda _{sdl}^{F}$ for permalloy our expressions for $\chi $ are not
applicable.

By taking into account the resistance of the interfaces between two FM
layers and the normal metal spacer, $AR_{I}$, we obtain 
\begin{equation}
\chi =\frac{1}{\lambda (1+r)}-1+\frac{r}{(1+r)(1-\gamma \gamma ^{\prime })},
\end{equation}
where $r=AR_{I}e^{2}N_{0}^{I}(\epsilon _{F})(1-\gamma ^{\prime \prime
2})/(1-\gamma \gamma ^{\prime })$, $e$ is the electron charge, $N_{0}^{I}$
is the density of states at the interface, $\gamma $, $\gamma ^{\prime }$, $
\gamma ^{\prime \prime }$ are the spin polarization parameters for the
conductivity, diffusion constant, and density of states at the interface
(see Appendix 2). We estimate $\chi $ to be about 31 for cobalt. 

\section{\label{sec6} Discussion of Results}

The salient conclusion we arrive at is that the bare currents do not
correctly estimate the amount of spin angular momentum transferred from the
polarized current to the background magnetization of the magnetic layers, in
the layered structures that have been studied to date. It is necessary to do
a complete ``globally diffusive'' transport calculation, with the
possibility of interfacial ballistic inserts to account for the spin
transfer there, in order to ascertain the enhancement of this spin transfer
by the accumulation attendant to CPP transport. The size of the spin
transfer region has not been resolved, but we can circumvent this
uncertainty by postulating a region $d_{\text{ }}$in which a transverse
component of the accumulation and current exist, and we can place in this
sector either results obtained from ab-initio 
calculation~\cite{others}$^-$\cite{StZ2}, 
or
our diffusive spin transfer$.$ Also when we consider diffuse interface
scattering the mean free path in the region of the spin transfer
is considerably smaller, by at least one order of magnitude, than in the
bulk of the layers.

The uncertainty in the size of the spin transfer region comes from
estimating the magnitude of the ``sd'' exchange interaction $J$. If one
erroneously identifies it with the spin splitting, $\Delta \thicksim 1$
eV found from band structure calculations which are limited to the
diagonal spin components of the exchange-correlation potential, one
would indeed find a spin transfer region no larger than $d_{k}$; however
it has been stressed that $J\thicksim kT_{C}\thicksim 0.1$ eV should be
identified as the ``Heisenberg-like'' exchange coupling found in
calculations that include the off-diagonal spin components of the
exchange-correlation potential~\cite {kubler}. The one case in which one
has been able to directly measure $J$ from transmission conduction
electron spin resonance one found $J=0.106$ eV (see Appendix~\ref{app1})
for permalloy~\cite{cooper}; unfortunately no data exists for
Co~\cite{hurdequint}. Indeed when we use this value for $J$ and the
$\lambda _{mfp}$ in the bulk of Co we find $\lambda _{J}\sim 3$ nm in
which case the spin transfer region would be larger than the $d_{k}$ of
the order of several angstroms anticipated by
others~\cite{others}$^-$\cite{StZ2}.  However, when we use the $\lambda
_{mfp}$\ appropriate to the interface between Co and Cu~\cite {bass}, we
find $\lambda _{J}\sim 1$ nm which is comparable to the upper limit
estimated from data on Co/Cu pillars~\cite{buhrman}.

Our conclusion about the amplification of the spin torque is independent of
the size of the spin transfer region (as long as $d$ is large enough so
that we can consider the conduction in the semi-classical approximation)
because our overall calculation of the diffusive transport outside the spin
transfer regions remains valid, inasmuch as it is identical to the well
established theory of Valet and Fert for CPP transport. However the {\it 
magnitude } of the amplification of the spin torque does depends on $d;$ see
Eq.~(\ref{aa}). {\em \ }The size of the spin torque transmitted to the
background is primarily governed by: the spin dependent transport parameters
of the thick magnetic layer which creates the polarized spin current, $\beta 
$, $\lambda _{sdl}^{F}$, the spin dependent interface scattering parameter $
\gamma $, and the resistivity and spin diffusion length in the normal 
back
layer relative to that of the thick magnetic layer. The characteristics for
the relatively thin magnetic layers $t_{F}$ used to observe current induced
switching do not determine the overall spin accumulation and current in the
sample; other than sensing the spin polarized current through the ``sd''
exchange interaction they do not affect the size of the spin torque acting
on the thin layer.

In our treatment of current induced switching we considered spin transport
across the entire CPP structure rather than for the interface region alone,
i.e., we have considered the spin torque due to the bulk of the magnetic
layers and that arising from diffuse scattering at interfaces. We have not
considered the problem of matching the distribution functions across
adjacent layers in the presence of {\it specular }scattering at the
interface; this requires a knowledge of the band structure in these layers
and is outside the scope of our study. The parameters entering our theory
are determined from CPP transport measurements, except for $J$, the "sd"
exchange interaction. Previous treatments highlighted the spin torque that
is attendant to ballistic transmission across an interface between magnetic
and nonmagnetic layers; as is the case for GMR reality is probably a mixture
of these two different positions.

\begin{acknowledgments}

We would like to thank Albert Fert, Herve Hurdequint, Jacques Miltat,
Charles Sommers and Mark Stiles for numerous helpful discussions; AS and
PML gratefully acknowledge the hospitality of the Laboratoire de
Physique des Solides at the Universit\'{e} Paris-Sud in Orsay, France
during a sabbatical leave for PML. This work was supported by the
National Science Foundation (DMR 0076171 and DMR 0131883), and the DoD
Multidisciplinary University Research Initiative (MURI) program
administered by the Office of Naval Research under Grant
N00014-96-1-1207, and the Defense Advanced Research Projects Agency
Contract No. MDA972-99-C-0009.

\end{acknowledgments}

\appendix

\section{}
\label{app1}

We separate the spin accumulation into longitudinal (parallel to the {\it 
local} moment) and transverse (perpendicular to the{\it \ local} moment)
modes. Equation (\ref{e}) can now be written as 
\begin{equation}
\frac{\partial ^{2}{\bf m}_{||}}{\partial x^{2}}-\frac{{\bf m}_{||}}{\lambda
_{sdl}^{2}}=\frac{j_{e}}{2D_{l}}\frac{\partial {\bf p}}{\partial x}\cdot 
{\bf M}_{d},  \label{f}
\end{equation}
where $\lambda _{sdl}=\sqrt{1-\beta \beta ^{\prime }}\lambda _{sf},D_{l}=
\sqrt{1-\beta \beta ^{\prime }}D_{0}$ and ${\bf p}(x)=\beta {\bf M}_{d}$,
which represents the {\it bare }spin polarization of the current coming
solely from the electric current in the absence of spin accumulation (see
Eqs.~(\ref{h}) and~(\ref{c})), and 
\begin{widetext}
\begin{equation}
\frac{\partial ^{2}{\bf m}_{\perp }}{\partial x^{2}}-\frac{{\bf m}_{\perp }}{
\lambda _{sf}^{2}}-\frac{{\bf m}_{\perp }\times {\bf M}_{d}}{\lambda _{J}^{2}
}=\frac{j_{e}}{2D_{0}}\left[ \frac{\partial {\bf p}}{\partial x}\times {\bf M
}_{d}-\left( \frac{\partial {\bf p}}{\partial x}\times {\bf M}_{d}\right)
\times {\bf M}_{d}\right] .  \label{g}
\end{equation}
\end{widetext}
The longitudinal accumulation ${\bf m}_{||}$ decays at the length scale of
the spin diffusion length $\lambda _{sdl}$ while the transverse spin
accumulation ${\bf m}_{\perp }$ decays as $\lambda _{J}$. In a typical
transition-metal ferromagnet, e.g. Co, the spin diffusion length $\lambda
_{sdl}$ has been measured to be about 60 nm~\cite{bass}. We 
estimate $
\lambda _{J}$ by taking the typical diffusion constant of a metal to be $
3\cdot 10^{-3}$ m$^{2}$/s and $J=0.1-0.4$ eV~\cite{hurdequint} so that 
$
\lambda _{J}$ is about 1.5 nm to 3 nm. Thus, the transverse spin
accumulation has a much shorter length scale compared to the longitudinal
one;{\em \ }it is larger than the mean free path in the interfacial region
between Co and Cu, $\thicksim 1$ nm, and is comparable to $\lambda _{mfp}\sim
6$\ nm in the bulk of Co. For permalloy, where we can use the value of $
J=0.106$ eV measured by conduction electron resonance~\cite{cooper} 
$\lambda
_{J}\sim 3$ nm which is comparable to $\lambda _{mfp}$ and $\lambda _{sdl}$
for permalloy. Therefore for multilayers containing Py our treatment is
not directly applicable as we assume in most of our work $\lambda _{J}\ll
\lambda _{sdl}$ ; we have to go back to the diffusion equation in 
Sec.~\ref{sec2} and solve the equations in this limit.

It makes a difference whether one treats the magnetization ${\bf M}_{d}(x)$
as a continuous function or as a finite difference. For example, in a domain
wall, where one treats the magnetization as a continuously rotating vector,
there is no longitudinal component of the spin accumulation ${\bf m}_{||}$
coming from the interior of the wall itself because $\frac{\partial }{
\partial x}{\bf p\thicksim }\frac{\partial }{\partial x}({\bf M}_{d})$ is
perpendicular, ``tangential'', to ${\bf M}_{d}$. In this case the transverse
component of the source term exists. In our treatment of transport across a
domain wall we accounted for the continuously rotating magnetization in the
wall by determining the correction to the electron states induced by the
rotations in spin space~\cite{levy}; we did not consider any spin
accumulation. Another treatment of the same problem by 
Simanek~\cite{simanek}
took an approach for the domain wall which is more consonant with the
equation of motion method we follow in this paper. In that approach the
transverse spin accumulation due to the continuously rotating magnetization
was determined, and he was able to calculate its contribution to the domain
wall resistance.

This continuous treatment for domain walls has to be contrasted with the
conventional treatment for current perpendicular to the plane of the layers
(CPP) in magnetic multilayers where one focuses on the discontinuous
variation of the magnetization between the layers; in this case one indeed
does have a longitudinal source term for the spin accumulation as we now
show. In the usual treatment of CPP we assume that the magnetization is
uniform throughout a layer so that the source term is confined to interfaces
between layers~\cite{zhang3},~\cite{valet-fert}; in this case one can 
take
into account the source terms by appropriate boundary conditions; this is
the procedure usually followed when calculating the spin accumulation in
magnetic multilayers~\cite{valet-fert}. By discretizing the source terms 
in
Eqs.~(\ref{f}) and~(\ref{g}) we find for the longitudinal accumulation 
\begin{equation}
\frac{\beta j_{e}}{2D_{0}}\sum_{j=i\pm 1}\hat{M}_{i}(1-\hat{M}_{i}\cdot \hat{
M}_{j})\delta (x_{j}),  \label{i}
\end{equation}
while for the transverse accumulation \ the source term is 
\begin{equation}
\frac{\beta j_{e}}{2D_{0}}\sum_{j=i\pm 1}\hat{M}_{i}\times (\hat{M}
_{i}\times \hat{M}_{j})\delta (x_{j}),  \label{j}
\end{equation}
where the layer we are considering is labelled $i$ while the interfaces with
the adjacent layers $j=i\pm 1$ are at $x_{j}$ . For {\it collinear }
structures we see that the transverse term is zero; the longitudinal source
term at the interface between two identical magnetic layers is zero if they
are parallel, and

\begin{equation}
\frac{\beta j_{e}}{D_{0}}\hat{M}_{i}\delta (x_{0})  \label{k}
\end{equation}
if they are antiparallel; here $x_{0}$ is the coordinate of the interface
between the two magnetic layers. At the interface between magnetic and
nonmagnetic layers (FM/NM) only the longitudinal source term exists, {\it 
irrespective of the alignment of neighboring magnetic layers}; it is 
\begin{equation}
\frac{\beta j_{e}}{2D_{0}}\hat{M}_{i}\delta (x_{0}).  \label{v}
\end{equation}
When two identical magnetic layers are noncollinear there is a transverse
source term given by Eq.~(\ref{j}) as well as a longitudinal one 
Eq.~(\ref{i}).

For the multilayered structure depicted in Fig.~\ref{fig1}, 
which models the case
studied up till now for current induced switching, no two magnetic layers
are adjacent so the sole source term that exists is given by 
Eq.~(\ref{i}).
In this case the boundary condition at the interfaces between adjacent FM/NM
layers is given by Eq.~(\ref{v}). However, as we make the assumption 
that the
thickness of the nonmagnetic spacer layer between the two magnetic layers is
much smaller than $\lambda _{sdl}^{N}$ we can replace the two sets of FM/NM
boundaries by one and use the conditions Eqs.~(\ref{i}) and~(\ref{j}).

\section{}
\label{app2}

\begin{figure}
\includegraphics[width=\textwidth]{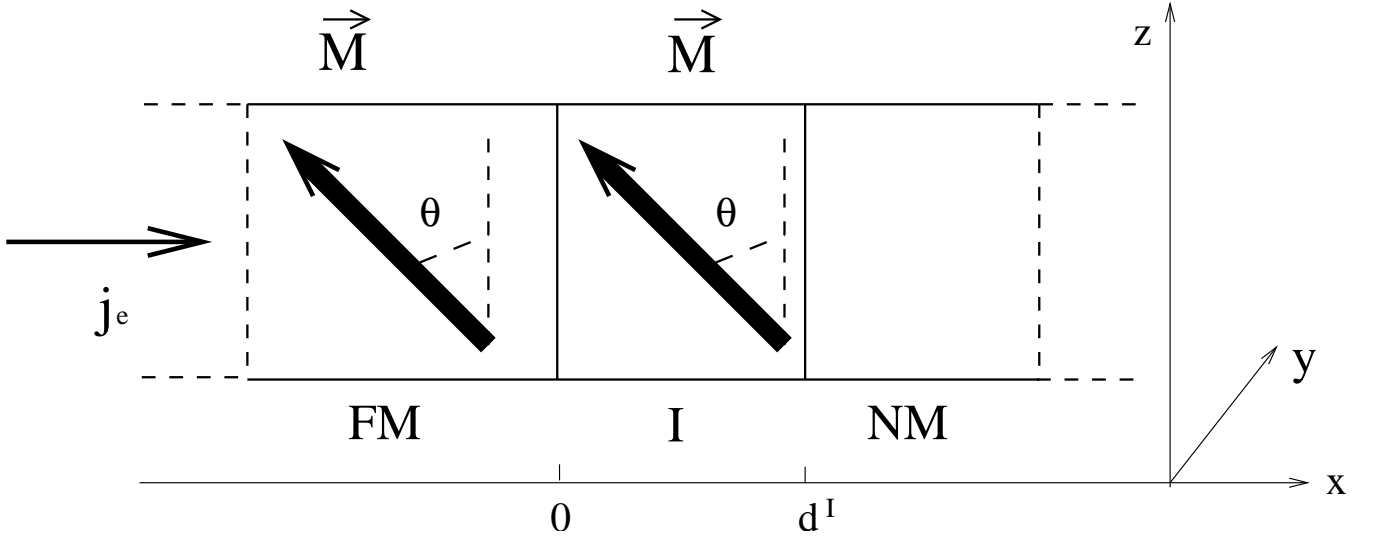}
\caption{Structure of the interface between the layers at Fig.1. FM is a
semi-infinite
ferromagnetic layer with the local magnetization ${\bf
M}_d^{(2)}=\cos\theta{\bf e}_z-\sin\theta{\bf e}_y$, I is a diffuse
interfacial layer with the same local magnetization as in FM layer, and
NM is a semi-infinite nonmagnetic layer.}
\label{fig6}
\end{figure}
In 
this Appendix, we derive the boundary conditions at the interfaces
between the layers in Fig.~\ref{fig1}. To achieve this goal, we consider
a sub-system shown in Fig.~\ref{fig6} which consists of a semi-infinite
FM layer with $x<0$ with the local magnetization ${\bf M=\cos \theta
e_{z}-\sin \theta e_{y}}$, a diffuse interfacial layer I between
$0<x<d^I$ with the same local magnetization as in the FM layer, and a
semi-infinite NM layer for $x>d^I$. When $d^I$ is infinitesimally small,
this sub-system represents the three FM-NM interfaces in
Fig.~\ref{fig1}, i.e., between the thick FM and spacer layers, between
the spacer and thin FM layers (when spatially inverted), and between the
thin FM and back NM layers. We assume that both spin accumulation and
current are continuous at the FM-I and I-NM interfaces, and derive the
relation between spin accumulation and current at $x=0$ with the same
quantities at $x=d^I$ as the thickness of the interfacial layer $d^I$
goes to zero. In this limit the parameters of the interfacial layer,
such as $\lambda _{mfp}^{I}$, $\tau _{sf}^{I}$, $J^{I}$, $\lambda
_{J}^{I}$, and, most important, its resistance $AR_{I}$ remain constant;
the latter condition implies that the diffusion constant of the
interfacial layer $D_{0}^{I}\thicksim d^I$ as $d^I\rightarrow 0$.

We solve Eqs.~(\ref{p}),~(\ref{q}) for the spin accumulation and use 
Eq.~(\ref{h}) 
to find spin current in the ferromagnetic and interfacial layers. By
adopting a set of local coordinates $(\bar{x},\bar{y},\bar{z})$ such 
that
the {\it local} magnetization is ${\bf M}_{\bar{z}}={\bf e}_{\bar{z}}$ 
the
spin accumulation and current in the FM layer take the form:
\begin{equation}
\left\{
\begin{array}{l}
m^F_{\bar{x}}=2Re\left( G_{2}\exp (\frac{x}{l_{+}^{F}})\right) \\
\\
m^F_{\bar{y}}=2Im\left( G_{2}\exp (\frac{x}{l_{+}^{F}})\right) \\
\\
m^F_{\bar{z}}=G_{1}\exp (\frac{x}{\lambda _{sdl}^{F}}),
\end{array}
\right.
\end{equation}
and
\begin{equation}
\left\{
\begin{array}{l}
j_{m,\bar{x}}^F=-4D_{0}^{F}Re\left( \frac{G_{2}}{l_{+}^{F}}\exp 
(\frac{x}{l_{+}^{F}})\right) \\
\\
j_{m,\bar{y}}^F=-4D_{0}^{F}Im\left( \frac{G_{2}}{l_{+}^{F}}\exp 
(\frac{x}{l_{+}^{F}})\right) \\
\\
j_{m,\bar{z}}^F=\beta j_{e}-\frac{2D_{0}^{F}(1-\beta \beta ^{\prime 
})}{\lambda _{sdl}^{F}}G_{1}\exp (\frac{x}{\lambda _{sdl}^{F}}).
\end{array}
\right.
\end{equation}
In the interfacial layer, spin accumulation and spin current take the 
form:
\begin{equation}
\left\{
\begin{array}{l}
m_{\bar{x}}^I=2Re\left( G_{5}\exp (\frac{x}{l_{+}^{I}})\right) 
+2Re\left(
G_{6}\exp (-\frac{x}{l_{+}^{I}})\right) \\
\\
m_{\bar{y}}^I=2Im\left( G_{5}\exp (\frac{x}{l_{+}^{I}})\right)+2Im\left(
G_{6}\exp (-\frac{x}{l_{+}^{I}})\right) \\
\\
m_{\bar{z}}^I=G_{3}\exp (\frac{x}{\lambda _{sdl}^{I}})+G_{4}\exp 
(-\frac{x}{\lambda_{sdl}^{I}}),
\end{array}
\right.
\end{equation}
\begin{widetext}
and
\begin{equation}
\left\{
\begin{array}{l}
j_{m,\bar{x}}^I=-4D_{0}^{I}\left[ Re\left( \frac{G_{5}}{l_{+}^{I}}\exp 
(\frac{x
}{l_{+}^{I}})\right) -Re\left( \frac{G_{6}}{l_{+}^{I}}\exp (-\frac{x}{
l_{+}^{I}})\right) \right] \\
\\
j_{m,\bar{y}}^I=-4D_{0}^{I}\left[ Im\left( \frac{G_{5}}{l_{+}^{I}}\exp 
(\frac{x
}{l_{+}^{I}})\right) -Im\left( \frac{G_{6}}{l_{+}^{I}}\exp (-\frac{x}{
l_{+}^{I}})\right) \right] \\
\\
j_{m,\bar{z}}^I=\gamma j_{e}-\frac{2D_{0}^{I}(1-\gamma \gamma ^{\prime 
})}{
\lambda _{sdl}^{I}}\left[ G_{3}\exp (\frac{x}{\lambda 
_{sdl}^{I}})-G_{4}\exp
(-\frac{x}{\lambda _{sdl}^{I}})\right] .
\end{array}
\right.
\end{equation}
Here $\beta $, $\beta ^{\prime }$, are spin-polarization 
parameters for 
the
conductivity defined in Sec.~\ref{sec2}; $\gamma $, $\gamma ^{\prime}$ are 
similar
parameters for the diffusion constant defined as ${\bf C}_{I}=\gamma
C_{0}^{I}{\bf M}_{d}$ in the bulk of the ferromagnetic layer, and ${\bf 
D}
_{I}=\gamma ^{\prime }D_{0}^{I}{\bf M}_{d}$ for the interfacial layer, 
and $
(l_{+}^{F,I})^{-1}=\sqrt{\frac{1}{(\lambda _{sf}^{F,I})^{2}}-\frac{i}{
(\lambda _{J}^{F,I})^{2}}}\approx \frac{1-i}{\sqrt{2}\lambda 
_{J}^{F,I}}$
when $\lambda _{J}^{F,I}\ll \lambda _{sf}^{F,I}$.

The boundary conditions for the continuity of the spin accumulation and
current at the interface between ferromagnetic and interfacial layer $x=0$
take the form:
\begin{equation}
\left\{
\begin{array}{l}
\label{m0}
2ReG_{2}=2ReG_{5}+2ReG_{6} \\
\\
2ImG_{2}=2ImG_{5}+2ImG_{6}) \\
\\
G_{1}=G_{3}+G_{4},
\end{array}
\right.
\end{equation}
and
\begin{equation}
\left\{
\begin{array}{l}
\label{j0}
-4D_{0}^{F}Re(\frac{G_{2}}{l_{+}^{F}}
)=-4D_{0}^{I}Re(\frac{G_{5}-G_{6}}{l_{+}^{I}}) \\
\\
-4D_{0}^{F}Im(\frac{G_{2}}{l_{+}^{F}})=
-4D_{0}^{I}Im(\frac{
G_{5}-G_{6}}{l_{+}^{I}}) \\
\\
\beta j_{e}-\frac{2D_{0}^{F}(1-\beta \beta ^{\prime 
})}{
\lambda _{sdl}^{F}}G_{1}=\gamma j_{e}-\frac{2D_{0}^{I}(1-\gamma \gamma
^{\prime })}{\lambda _{sdl}^{I}}(G_{3}-G_{4}).
\end{array}
\right.
\end{equation}

To relate ${\bf m}^{F}(0)$ to ${\bf m}^{I}(d^I)$, and ${\bf j}_m^F(0)$ 
to ${\bf j}_m^I(d^I)$, we use the assumption that as the thickness of 
the
interfacial layer goes to zero, other parameters of the interfacial 
layer,
such as $\lambda _{sdl}^{I}$, $J^{I}$, and $\lambda _{J}^{I}$ remain
constant, but the diffusion constant $D_{0}^{I}$ goes to zero with the 
same
rate as $d^I$, so that $d^I/D_{0}^{I}=const$. Then, for example, for 
small $d^I\ll \lambda_{J}^{I}$ the $\bar{x}$-component of the 
spin-accumulation at $x=d^I$
may be written as
\begin{equation}
m_{\bar{x}}^I(d^I\rightarrow 0)\approx 2Re(G_{5}+G_{6})+2Re\left(
(G_{5}-G_{6})\frac{d^I}{l_{+}^{I}}\right) .  \label{bb}
\end{equation}
By comparing this expression with Eqs.~(\ref{m0}) and~(\ref{j0}), we 
obtain
a relation between the $\bar{x}$-components of spin accumulation and 
current
at $x=0$ and $x=d^I$:
\begin{equation}
m_{\bar{x}}^I(d^I\rightarrow 
0)=m_{\bar{x}}^F(0)-j_{m,\bar{x}}^F(0)\frac{d^I}{
2D_{0}^{I}},  \label{mx}
\end{equation}
and similarly,
\begin{equation}
m_{\bar{y}}^I(d^I\rightarrow 
0)=m_{\bar{y}}^F(0)-j_{m,\bar{y}}^F(0)\frac{d^I}{
2D_{0}^{I}},
\end{equation}
\begin{equation}
m_{\bar{z}}^I(d^I\rightarrow 0)=m_{\bar{z}}^F(0)+j_{e}\frac{\gamma }{
2(1-\gamma \gamma ^{\prime 
})}\frac{d^I}{D_{0}^{I}}-j_{m,\bar{z}}^F(0)\frac{1}{
2(1-\gamma \gamma ^{\prime })}\frac{d^I}{D_{0}^{I}}.
\end{equation}
In a manner similar to Eq.~(\ref{bb}) the $\bar{x}$-component of spin 
current
at $x=d^I$ may be written as
\[
j_{m,\bar{x}}^I(d^I\rightarrow 0)\approx -4D_{0}^{I}Re\left( 
\frac{G_{5}-G_{6}
}{l_{+}^{I}}\right) -2Re\left( i(G_{5}+G_{6})\right) 
\frac{d^{I}J^{I}}{\hbar }.
\]
By comparing this expression with Eqs.~(\ref{m0}) and~(\ref{j0}), we 
find
the continuity condition for the $\bar{x}$-component of spin \ 
current:
\begin{equation}
j_{m,\bar{x}}^I(d^I\rightarrow 
0)=j_{m,\bar{x}}^F(0)-m_{\bar{y}}^F(0)\frac{
d^{I}J^{I}}{\hbar },
\end{equation}
and, similarly,
\begin{equation}
j_{m,\bar{y}}^I(d^I\rightarrow 
0)=j_{m,\bar{y}}^F(0)+m_{\bar{x}}^F(0)\frac{
d^{I}J^{I}}{\hbar },
\end{equation}
and
\begin{equation}
j_{m,bar{z}}^I(d^I\rightarrow 
0)=j_{m,\bar{z}}^F(0)-m_{\bar{z}}^F(0)\frac{d^I}{
\tau_{sf}^{I}}.  \label{jz}
\end{equation}
With these relations we can now obtain the boundary conditions at the 
three
interfaces in the multilayered structure depicted in Fig.~\ref{fig1}.

By using the conditions Eqs.~(\ref{mx})-(\ref{jz}), the boundary 
conditions
at the interface between the thin (first) ferromagnetic and non-magnetic (N)
layers of the structure shown in Fig.~\ref{fig1} at $x=t_{F}$ may be written
immediately, since in the thin FM layer the local coordinate system 
$(\bar{x}
,\bar{y},\bar{z})$ coincides with the global axes $(x,y,z)$; we find
\begin{equation}
\left\{
\begin{array}{l}
\label{bctFm} 
m^N_x(t_{F})-m^{(1)}_x(t_{F})=-rj^{(1)}_{m,x}(t_{F}) \\
\\
m^N_y(t_{F})-m^{(1)}_y(t_{F})=-rj^{(1)}_{m,y}(t_{F}) \\
\\
m^N_z(t_{F})-m^{(1)}_z(t_{F})=rj_{e}\frac{\gamma }{1-\gamma \gamma 
^{\prime }}
-rj^{(1)}_{m,z}(t_{F})\frac{1}{1-\gamma\gamma^{\prime }},
\end{array}
\right.
\end{equation}
and
\begin{equation}
\left\{
\begin{array}{l}
\label{bctFj} 
j^N_{m,x}(t_F)-j^{(1)}_{m,x}(t_{F})=
-m^{(1)}_{y}(t_F)\frac{d^{I}J^I}{\hbar}\\
\\
j^N_{m,y}(t_{F})-j^{(1)}_{m,y}(t_{F})=
m^{(1)}_{x}(t_{F})\frac{d^{I}J^I}{\hbar} 
\\
\\
j^N_{m,z}(t_{F})-j^{(1)}_{m,z}(t_{F})=
-m^{(1)}_{z}(t_{F})\frac{d^{I}}{\tau^I_{sf}},
\end{array}
\right.
\end{equation}
where $r=d^{I}/2D_{0}^{I}$. Similarly, the boundary conditions at the 
interface
between the non-magnetic spacer (S) and the thin (first) FM layer at 
$x=0$
take the form:
\begin{equation}
\left\{
\begin{array}{l}
\label{SpFMm} 
m^S_{x}(0)-m^{(1)}_{x}(0)=rj^{(1)}_{m,x}(0) \\
\\
m^S_{y}(0)-m^{(1)}_{y}(0)=rj^{(1)}_{m,y}(0) \\
\\
m^S_{z}(0)-m^{(1)}_{z}(0)=-rj_{e}\frac{\gamma }{1-\gamma \gamma 
^{\prime}}
+rj^{(1)}_{m,z}(0)\frac{1}{1-\gamma \gamma ^{\prime }},
\end{array}
\right.
\end{equation}
and
\begin{equation}
\left\{
\begin{array}{l}
\label{SpFMj} 
j^S_{m,x}(0)-j^{(1)}_{m,x}(0)=m^{(1)}_{y}(0)\frac{d^{I}J^I}{\hbar} \\
\\
j^S_{m,y}(0)-j^{(1)}_{m,y}(0)=-m^{(1)}_{x}(0)\frac{d^{I}J^I}{\hbar} \\
\\
j^S_{m,z}(0)-j^{(1)}_{m,z}(0)=m^{(1)}_{z}(0)\frac{d^{I}}{\tau^I_{sf}},
\end{array}
\right.
\end{equation}

Note that spin-current conservation condition at the interfaces, which 
means
that there are no torques acting at the interfaces, is due to the infinitely
small thickness of the interfacial layers $d^{I}\rightarrow 0$.{\em \ }
To write
the boundary conditions at the interface between the thick FM and NM 
spacer
layers at $x=0$, we have to change from the local coordinate system 
$(\bar{x}
,\bar{y},\bar{z})$, related to the magnetization direction in the thick 
FM
layer, to the global $(x,y,z)$ system. Any vector ${\bf a}$ will be
transformed according to the following rule:
\begin{equation}
\left\{
\begin{array}{l}
a_{\bar{x}}=a_{x} \\
\\
a_{\bar{y}}=a_{y}\cos \theta +a_{z}\sin \theta \\
\\
a_{\bar{z}}=-a_{y}\sin \theta +a_{z}\cos \theta .
\end{array}
\right.
\end{equation}
By applying this transformation to the conditions, 
Eqs.~(\ref{mx})-(\ref{jz}), 
we obtain the following boundary conditions at the interface between the
thick (second) FM and non-magnetic spacer (S) layers:
\begin{equation}
\left\{
\begin{array}{c}
\label{FMSpm}
m^S_{x}(0)-m^{(2)}_{x}(0)=-rj^{(2)}_{m,x}(0) \\
\\
m^S_{y}(0)-m^{(2)}_{y}(0)=
-rj_{e}\frac{\gamma }{1-\gamma \gamma ^{\prime}}\sin\theta 
-rj^{(2)}_{m,y}(0)\frac{1-\gamma \gamma ^{\prime }\cos ^{2}\theta }{
1-\gamma \gamma ^{\prime }} \\
\\
+rj^{(2)}_{m,z}(0)\sin \theta \cos \theta \frac{\gamma \gamma ^{\prime 
}}{1-\gamma\gamma ^{\prime }} \\
\\
m^S_{z}(0)-m^{(2)}_{z}(0)=rj_{e}\frac{\gamma }{1-\gamma \gamma ^{\prime 
}}\cos\theta 
+rj^{(2)}_{m,y}(0)\sin \theta \cos \theta \frac{\gamma \gamma ^{\prime 
}}{
1-\gamma \gamma ^{\prime }} \\
\\
+rj^{(2)}_{m,z}(0)\frac{1-\gamma \gamma ^{\prime }\sin ^{2}\theta 
}{1-\gamma\gamma ^{\prime }},
\end{array}
\right.
\end{equation}
and
\begin{equation}
\left\{
\begin{array}{c}
\label{FMSpj}
j^S_{m,x}(0)-j^{(2)}_{m,x}(0)=-m^{(2)}_{y}(0)\cos \theta 
\frac{d^{I}J^{I}}{\hbar
}-m^{(2)}_{z}(0)\sin \theta \frac{d^{I}J^{I}}{\hbar } \\
\\
j^S_{m,y}(0)-j^S_{m,y}(0)=m^{(2)}_{x}(0)\cos \theta 
\frac{d^{I}J^{I}}{\hbar }
-m^{(2)}_{y}(0)\sin ^{2}\theta \frac{d^{I}}{\tau _{sf}^{I}} \\
\\
+m^{(2)}_{z}(0)\sin \theta \cos \theta \frac{d^{I}}{\tau _{sf}^{I}} \\
\\
j^S_{m,z}(0)-j^{(2)}_{m,z}(0)=m^{(2)}_{x}(0)\sin \theta 
\frac{d^{I}J^{I}}{\hbar }
+m_{2y}(0)\sin \theta \cos \theta \frac{d^{I}}{\tau _{sf}^{I}} \\
\\
-m^{(2)}_{z}(0)\cos ^{2}\theta \frac{d^{I}}{\tau _{sf}^{I}},
\end{array}
\right.
\end{equation}

Note that as the thickness of the interfacial layer $d^{I}$ goes to zero, for
diffuse scattering we considered one produces large discontinuities in 
the
spin-accumulation (Eqs.~(\ref{bctFm}),~(\ref{SpFMm}),~(\ref{FMSpm})) 
proportional
to finite $r=d^{I}/2D_{0}^{I}$, but small discontinuities in the 
spin-currents
(Eqs.~(\ref{bctFj}),~(\ref{SpFMj}),~(\ref{FMSpj})) proportional to $d^{I}$, 
because 
$J_{I}$ does not increase and $\tau _{sf}^{I}$ does not decrease as 
$d^{I}\rightarrow 0$. In our picture finite thickness of the interfacial 
layer is
essential for torque production at the interface. As we consider 
infinitely
small interfacial thicknesses $d^{I}\rightarrow 0$, we obtain 
spin-current
conservation conditions at each interface:
\begin{equation}
{\bf j}^N_{m}(t_{F})={\bf j}^{(1)}_{m}(t_{F}),  \label{bctFjc}
\end{equation}
\begin{equation}
{\bf j}^S_{m}(0)={\bf j}^{(1)}_{m}(0),  \label{SpFMjc}
\end{equation}
and
\begin{equation}
{\bf j}^{(2)}_{m}(0)={\bf j}^S_{m}(0).  \label{FMSpjc}
\end{equation}

By eliminating $m^{S}(0)$ and $j^S_{m}(0)$ from Eqs.~(\ref{SpFMm}), 
(\ref
{FMSpm}), (\ref{SpFMjc}), and (\ref{FMSpjc}), we finally obtain the boundary
conditions at the interface between thick (second) and thin (first) FM
layers at $x=0$ (of course there is the NM spacer in-between, however 
its
thickness $t_{N}$ is irrelevant for these boundary conditions as long as 
$
t_{N}\ll \lambda _{sdl}^{N}$):
\begin{equation}
\left\{
\begin{array}{c}
\label{bc0m}
m^{(1)}_{x}(0)-m^{(2)}_{x}(0)=-2rj^{(1)}_{m,x}(0) \\
\\
m^{(1)}_{y}(0)-m^{(2)}_{y}(0)=
-rj_{e}\frac{\gamma }{1-\gamma \gamma ^{\prime }}\sin\theta 
-rj^{(1)}_{m,y}(0)\frac{2-\gamma \gamma ^{\prime }(1+\cos ^{2}\theta 
)}{
1-\gamma \gamma ^{\prime }} \\
\\
+rj^{(1)}_{m,z}(0)\sin \theta \cos \theta 
\frac{\gamma \gamma ^{\prime }}{1-\gamma\gamma ^{\prime }} \\
\\
m^{(1)}_{z}(0)-m^{(2)}_{z}(0)=
rj_{e}\frac{\gamma }{1-\gamma \gamma ^{\prime }}(1+\cos\theta )+
rj^{(1)}_{m,y}(0)\sin \theta \cos \theta \frac{\gamma \gamma ^{\prime 
}}{
1-\gamma \gamma ^{\prime }} \\
\\
-rj^{(1)}_{m,z}(0)\frac{2-\gamma \gamma ^{\prime }\sin ^{2}\theta 
}{1-\gamma
\gamma ^{\prime }},
\end{array}
\right.
\end{equation}
and
\begin{equation}
{\bf j}^{(1)}_{m}(0)={\bf j}^{(2)}_{m}(0)  \label{bc0j}
\end{equation}

Finally, we show that parameter $r=d^{I}/2D_{0}^{I}$ is proportional to 
the
interface resistance $AR_{I}$ found from CPP transport measurements 
~\cite
{basspratt}. By considering the expression (\ref{je}) for the electrical
current in the interfacial layer, and the assumptions that $
D_{0}^{I}\thicksim d^{I}$ and $\lambda _{sdl}^{I}$ remains constant as 
the
thickness of the interfacial layer $d^{I}\rightarrow 0$, we find 
$AR_{I}=d^{I}/2C_{0}^{I}$, or 
$r=\frac{d^{I}}{2D_{0}^{I}}=AR_{I}\frac{C_{0}^{I}}{
D_{0}^{I}}$. The parameters $C_{0}^{I}$ and $D_{0}^{I}$ may be related 
via
Einstein's relation $\hat{C}_{I}=e^{2}\hat{N}_{I}(\epsilon 
_{F})\hat{D}_{I}$
, so that the parameter $r$ takes the form:
\begin{equation}
r=AR_{I}e^{2}N_{0}^{I}(\epsilon _{F})\frac{1-\gamma ^{\prime \prime 
2}}{
1-\gamma \gamma ^{\prime }},  \label{res}
\end{equation}
where $e$ is the electron charge, $N_{0}^{I}(\epsilon _{F})$ is the 
density
of states at the interface at Fermi energy, and $\gamma ^{\prime \prime 
}$
is the spin polarization parameters for the density of states at the
interfaces which is defined as ${\bf N}_{I}=\gamma ^{\prime \prime 
}N_{0}^{I}
{\bf M}_{d}$.

\section{}
\label{app3}

We solve the Eqs.~(\ref{p}),~(\ref{q}) for the spin 
accumulation in each of
three layers, and find spin currents using 
Eq.~(\ref{h}). In the thick
ferromagnetic layer, spin accumulation and current take 
the form
\begin{equation}
\left\{
\begin{array}{l}
m^{(2)}_{x}=2Re\left( G_{2}\exp 
(\frac{x}{l_{+}})\right) \\
\\
m^{(2)}_{y}=2Im\left( G_{2}\exp 
(\frac{x}{l_{+}})\right) 
\cos \theta -G_{1}\exp (
\frac{x}{\lambda^F_{sdl}})\sin \theta \\
\\
m^{(2)}_{z}=2Im\left( G_{2}\exp 
(\frac{x}{l_{+}})\right) 
\sin \theta +G_{1}\exp (
\frac{x}{\lambda^F_{sdl}})\cos \theta ,
\end{array}
\right.
\end{equation}
and
\begin{equation}
\left\{
\begin{array}{c}
j^{(2)}_{m,x}=-4D_{0}Re\left( \frac{G_{2}}{l_{+}}
\exp (\frac{x}{l_{+}})\right) 
\\
\\
j^{(2)}_{m,y}=-\beta j_{e}\sin \theta -4D_{0}Im\left( 
\frac{G_{2}}{l_{+}}\exp (
\frac{x}{l_{+}})\right) \cos \theta \\
\\
+\frac{2D_{0}(1-\beta \beta ^{\prime })}{\lambda 
_{sdl}}G_{1}\exp (\frac{x}{
\lambda^F_{sdl}})\sin \theta \\
\\
j^{(2)}_{m,z}=\beta j_{e}\cos \theta -4D_{0}Im\left( 
\frac{G_{2}}{l_{+}}\exp (
\frac{x}{l_{+}})\right) \sin \theta 
-\frac{2D_{0}(1-\beta \beta ^{\prime })}{
\lambda^F_{sdl}}G_{1}\exp 
(\frac{x}{\lambda^F_{sdl}})\cos \theta .
\end{array}
\right.
\end{equation}
In the thin ferromagnetic layer
\begin{equation}
\left\{
\begin{array}{l}
m^{(1)}_{x}=2Re\left( G_{5}\exp 
(-\frac{x}{l_{+}})\right) 
+2Re\left( G_{6}\exp (
\frac{x-t_{F}}{l_{+}})\right) \\
\\
m^{(1)}_{y}=2Im\left( G_{5}\exp 
(-\frac{x}{l_{+}})\right) 
+2Im\left( G_{6}\exp (
\frac{x-t_{F}}{l_{+}})\right) \\
\\
m^{(1)}_{z}=G_{3}\exp (-\frac{x}{\lambda^F_{sdl}})+
G_{4}\exp (\frac{x-t_{F}}{\lambda^F_{sdl}}),
\end{array}
\right.
\end{equation}
and
\begin{equation}
\left\{
\begin{array}{l}
j^{(1)}_{m,x}=4D_{0}\left[ Re\left( 
\frac{G_{5}}{l_{+}}\exp 
(-\frac{x}{l_{+}}
)\right) -Re\left( \frac{G_{6}}{l_{+}}\exp 
(\frac{x-t_{F}}{l_{+}})\right)
\right] \\
\\
j^{(1)}_{m,y}=4D_{0}\left[ Im\left( 
\frac{G_{5}}{l_{+}}\exp 
(-\frac{x}{l_{+}}
)\right) -Im\left( \frac{G_{6}}{l_{+}}\exp 
(\frac{x-t_{F}}{l_{+}})\right)
\right] \\
\\
j^{(1)}_{m,z}=\beta j_{e}+\frac{2D_{0}(1-\beta \beta 
^{\prime })}
{\lambda^F_{sdl}}
\left[ G_{3}\exp (-\frac{x}{\lambda^F_{sdl}})-G_{4}\exp 
(\frac{x-t_{F}}{
\lambda^F_{sdl}})\right] .
\end{array}
\right.
\end{equation}
where $l_{+}^{-1}=\sqrt{\frac{1}{\lambda 
_{sf}^{2}}-\frac{i}{\lambda _{J}^{2}
}}\approx \frac{1-i}{\sqrt{2}\lambda _{J}}$, and 
$\lambda^F_{sdl}$ is
spin-diffusion length in FM layer. In the non-magnetic 
layer,
\begin{equation}
{\bf m}^{N}={\bf A}\exp (-\frac{x-t_{F}}{\lambda 
_{sdl}^{N}}),
\end{equation}
and
\begin{equation}
{\bf j}^N_{m}=\frac{2D_{0}^{N}}{\lambda _{sdl}^{N}}{\bf A}\exp 
(-\frac{x-t_{F}
}{\lambda _{sdl}^{N}}).
\end{equation}

To obtain the 12 unknown constants $A_{x}$, $A_{y}$, $A_{z}$, $G_{1}$, 
$ReG_{2}$, $ImG_{2}$, $G_{3}$, $G_{4}$, $ReG_{5}$, 
$ImG_{5}$, $ReG_{6}$, $
ImG_{6}$, we use the boundary conditions (see Appendix~\ref{app2}, Eqs.~(\ref{bctFm}
), (\ref{bc0m}), (\ref{bctFjc}), and (\ref{bc0j})):
\begin{equation}
\left\{
\begin{array}{l}
m^N_{x}(t_{F})-m^{(1)}_{x}(t_{F})=-rj^{(1)}_{m,x}(t_{F}) 
\\
\\
m^N_{y}(t_{F})-m^{(1)}_{y}(t_{F})=-rj^{(1)}_{m,y}(t_{F}) 
\\
\\
m^N_{z}(t_{F})-m^{(1)}_{z}(t_{F})=
rj_{e}\frac{\gamma }{1-\gamma \gamma ^{\prime }}
-rj^{(1)}_{m,z}(t_{F})\frac{1}{1-\gamma \gamma ^{\prime 
}},
\end{array}
\right.
\end{equation}
and
\begin{equation}
\left\{
\begin{array}{c}
m^{(1)}_{x}(0)-m^{(2)}_{x}(0)=-2rj^{(1)}_{m,x}(0) \\
\\
m^{(1)}_{y}(0)-m^{(2)}_{y}(0)=-rj_{e}\frac{\gamma 
}{1-\gamma \gamma 
^{\prime }}\sin\theta -rj^{(1)}_{m,y}(0)
\frac{2-\gamma \gamma ^{\prime }(1+\cos ^{2}\theta )}{
1-\gamma \gamma ^{\prime }} \\
\\
+rj^{(1)}_{m,z}(0)\sin \theta \cos \theta 
\frac{\gamma \gamma ^{\prime }}{1-\gamma
\gamma ^{\prime }} \\
\\
m^{(1)}_{z}(0)-m^{(2)}_{z}(0)=
rj_{e}\frac{\gamma }{1-\gamma \gamma ^{\prime 
}}(1+\cos\theta )
+rj^{(1)}_{m,y}(0)\sin \theta \cos \theta 
\frac{\gamma \gamma ^{\prime }}{1-\gamma \gamma 
^{\prime }} \\
\\
-rj^{(1)}_{m,z}(0)\frac{2-\gamma \gamma ^{\prime }
\sin ^{2}\theta }{1-\gamma\gamma ^{\prime }},
\end{array}
\right.   \label{w}
\end{equation}
\begin{equation}
{\bf j}^N_{m}(t_{F})={\bf j}^{(1)}_{m}(t_{F})
\end{equation}
\begin{equation}
{\bf j}^{(1)}_{m}(0)={\bf j}^{(2)}_{m}(0)
\end{equation}
where the parameter $r$ is proportional to the 
interface resistance $AR_{I}$
, $r=AR_{I}e^{2}N_{0}(1-\gamma ^{\prime \prime 
2})/(1-\gamma \gamma ^{\prime
})$, $e$ is the electron charge, $N_{0}$ is the density 
of states at the
interface, $\gamma $, $\gamma ^{\prime }$, $\gamma 
^{\prime \prime }$ are
the spin polarization parameters for the conductivity, 
diffusion constant,
and density of states at the interfaces (see Appendix~\ref{app2}). 
The other six boundary conditions come from the conservation of spin 
current at the interfaces.
\end{widetext}


\end{document}